\documentclass[a4paper, 10pt, twocolumn]{IEEEtran}
\usepackage[T1]{fontenc}
\usepackage[english]{babel}
\usepackage{verbatim}
\usepackage{amsmath}
\usepackage{cite}
\usepackage{graphics}
\usepackage{epsfig}
\usepackage{amssymb}
\PassOptionsToPackage{normalem}{ulem}
\usepackage{ulem}
\usepackage{subfigure}
\usepackage{caption}
\usepackage{color}
\providecommand{\tabularnewline}{\\}

\begin{document}

\title{A Particle Multi-Target Tracker for Superpositional Measurements using Labeled Random Finite Sets}

\author{Francesco Papi and Du Yong Kim
		\thanks{Copyright (c) 2015 IEEE. Personal use of this material is permitted.
				However, permission to use this material for any other purposes must
				be obtained from the IEEE by sending a request to pubs-permissions@ieee.org} 
		\thanks{Acknowledgement: This work is supported by the National Security and 
				Intelligence, Surveillance and Reconnaissance Division (NSID), Defence Science 
				and Technology Organisation (DSTO), Edinburgh, SA, 5111.}
		\thanks{Francesco Papi and Du Yong Kim are with the Department of Electrical 
				and Computer Engineering, Curtin University, Bentley,
				WA 6102, Australia. E-mail: \{francesco.papi, duyong.kim\}@curtin.edu.au}}

\maketitle		
		
\begin{abstract}
In this paper we present a general solution for multi-target tracking with \textit{superpositional measurements}. Measurements that are functions of the sum of the contributions of
the targets present in the surveillance area are called \textit{superpositional measurements}.  We base our modelling on Labeled
Random Finite Set (RFS) in order to jointly estimate the number of targets and their trajectories. 
This modelling leads to a labeled version of Mahler's multi-target Bayes filter.
However, a straightforward implementation of this tracker using Sequential Monte Carlo (SMC) methods is not feasible due to 
the difficulties of sampling in high dimensional spaces.
We propose an efficient multi-target sampling strategy based on Superpositional Approximate CPHD (SA-CPHD) filter and the recently introduced Labeled Multi-Bernoulli (LMB)
and Vo-Vo densities. The applicability of the proposed approach is verified through simulation in a challenging radar 
application with closely spaced targets and low signal-to-noise ratio.
\end{abstract}


\section{INTRODUCTION}

\emph{Superpositional sensors} are an important class of pre-detection
sensor models which arise in a wide range of joint detection and estimation problems. 
For example, in problems such as direction-of-arrival estimation for linear antenna arrays \cite{Balakumar05},
multi-user detection for wireless communication networks \cite{Angelosante07},
acoustic amplitude sensors \cite{Hlinka2012}, radio frequency (RF) tomography \cite{Nan13}, 
target tracking with unresolved or merged measurements \cite{Svensson2012,Beard2015},
multi-target Track-Before-Detect with closely spaced targets \cite{DaveyTBDbook,PapGLMB14},
the sensor output is a function of the sum of contributions from individual sources. In classical estimation theory, a frequency domain model for the superpositional sensor is generally used to design algorithms for source separation and parameter estimation. Conversely in dynamic state estimation, a detection based model is typically employed to transform the collected data into a set of point measurements, in order to facilitate the development of computationally efficient estimation algorithms. This is specifically the case in multi-target tracking \cite{Blackman1999,Mahler2007,Bar2011}, which is an important problem in estimation theory involving
the joint estimation of an unknown and time varying number of targets and their trajectories.


Many real life applications in radar, sonar \cite{Blackman1999,Mahler2007,Bar2011,battistelli2012robust}, 
computer vision \cite{Woo2007,bramble,HVV_TSP_13}, robotics \cite{Montemerlo2007,
Lee2013,Durrantwhyte2006,Mullane2011}, automotive safety \cite{Battistelli,Meisner}, 
cell biology \cite{Ray2002,Smal2008,Huang2009, Chat2013,Rezatofighi2015}, etc., 
can be described as multi-target tracking problems.
Most of the multi-target tracking algorithms existing in the literature are 
designed for data that have been preprocessed into point measurements or detections 
\cite{Blackman1999,Mahler2007,Bar2011,battistelli2011optimal}. 
These algorithms are based on the ``detection sensor'' model which
assumes that each target generates at most one detection, and that
each measurement belongs to at most one target \cite{Mahler2007}.
The performed preprocessing of raw measurements into a finite sets of points is
efficient in terms of memory and computational requirements, and is
usually effective for a wide range of applications. 
However, the compression might lead to significant information loss in the
presence of low signal-to-noise ratio (SNR) and/or closely spaced targets.
The standard ``detection sensor'' approach may not be adequate in this case,
and making use of all information contained in the pre-detection measurements
becomes necessary. In turn, this requires more advanced sensor models and new algorithms.

In a superpositional sensor model, the measurement at each time step
is a superposition of measurements generated by each of the targets
present \cite{MahSCPHD}. 
In \cite{MahSCPHD} Mahler derived a superpositional Cardinalized Probability Hypthesis Density (CPHD) filter as
a tractable approximation to the Bayes multi-target filter for superpositional sensor. 
The approach was implemented in \cite{Nan13} using SMC methods, and successfully applied to a passive acoustics 
application as well as RF tomography. The technique was also extended to multi-Bernoulli and a combination of 
multi-Bernouli and CPHD \cite{Nan13a,Nan14}.
These filters, however, are not multi-target trackers because they
rest on the premise that targets are indistinguishable. Moreover,
they require at least two levels of approximations: analytic approximations
of the Bayes multi-target filter and particles approximation of the obtained recursion.

Inspired by \cite{MahSCPHD,Nan13}, this paper proposes a multi-target
tracker for superpositional sensors which estimates target tracks and
requires only one level of approximation. Our formulation is based
on the same random finite set (RFS) framework that the superpositional
CPHD filters \cite{MahSCPHD,Nan13} were derived from. However, we
used a special class of RFS models, called labelled RFS \cite{BTV13},
which enables the estimation of target tracks as well as direct particle
approximation of the (labeled) Bayes multi-target filter. To mitigate
the depletion problem arising from sampling in high dimensional space
we propose an efficient multi-target sampling strategy using the superpositional
CPHD filter \cite{Nan13}. 
In particular, we will show how the recently introduced Labeled Multi-Bernoulli \cite{Reuter2013}
and Vo-Vo \footnote{The Vo-Vo density was originally called the Generalized 
Labeled Multi-Bernoulli density. However, for compactness we follow Mahler's
latest book \cite{Mahler2014} and call this the Vo-Vo density} 
\cite{Vo2014} densities can be constructed from the superpositional approximate 
CPHD (SA-CPHD) filter. These densities are then used to design effective proposal distributions 
for the RFS multi-target particle filter.
While both the CPHD and labeled RFS solutions
require particle approximation, the latter has the advantage that
it does not require particle clustering for the multi-target state
estimation. The applicability of the proposed approach is verified
through simulation analyses in a challenging closely-spaced multi-target 
scenario using radar power measurements with low signal-to-noise ratio (SNR) \cite{Boe04}.

The paper is organised as follows: in Section \ref{sec:Background} we recall 
some definitions for Labeled RFSs and superpositional sensors. In Section 
\ref{sec:Multitarget_tracker} we discuss the multi-target particle filter,
the labeled multi-target transition density, and the superpositional approximate 
CPHD. Two multi-target particle trackers using the Labeled Multi-Bernoulli (LMB)
and the Vo-Vo  densities for 
the proposal distribution are presented in Section \ref{sec:Proposals}.
Numerical results for a radar application are presented in Section \ref{sec:Simulations},
while conclusions and future research directions are discussed in Section \ref{sec:Conclusion}.

\section{Background\label{sec:Background}}

This section briefly presents background material on superpositional
sensor model and the RFS framework which we adopt for the formulation of
 a multi-target tracking filter. Subsection \ref{sub:Multi-target-Estimation} provides
a summary of basic concepts in RFS. We present a concise description
of the superpositional sensor model in subsection \ref{sub:Superpositional-Sensor}
and report a summary of key ideas on labled RFS needed for the derivation
in subsection \ref{sub:Labeled-RFS}.

\subsection{Multi-target Estimation\label{sub:Multi-target-Estimation}}

Suppose that at time $k$, there are $N(k)$ target states $x_{k,1},\ldots,x_{k,N(k)}$,
each taking values in a state space $\mathcal{X}$. In the random
finite set (RFS) framework, the \emph{multi-target state} at time
$k$ is represented by the finite set $X_{k}=\{x_{k,1},\ldots,x_{k,N(k)}\}$,
and the multi-target state space is the space of all finite subsets
of $\mathcal{X}$, denoted as $\mathcal{F}\mathbf{(}\mathcal{X}\mathbf{)}$.
An RFS is simply a random variable that take values the space $\mathcal{F}\mathbf{(}\mathcal{X}\mathbf{)}$
that does not inherit the usual Euclidean notion of integration and
density. Mahler's Finite Set Statistics (FISST) provides powerful
yet practical mathematical tools for dealing with RFSs \cite{MahlerPHD2,Mahler2007}
based on a notion of integration/density that is consistent with point
process theory \cite{VSD05}.

Similar to the standard state space model, the multi-target system
model can be specified, for each time step $k$, via the \emph{multi-target
transition density} $f_{k|k-1\!}$ and the \emph{multi-target likelihood}
\emph{function} $g_{k}$, using the FISST notion of integration/density.
The \emph{multi-target posterior density }(or simply multi-target
posterior) contains all information about the multi-target states
given the measurement history. The multi-target posterior recursion
is direct generalisation of the standard posterior recursion \cite{Doucet2000},
i.e.
\begin{eqnarray}
\pi_{0:k}(X_{0:k}|z_{1:k})\propto\ \ \ \ \ \ \ \ \ \ \ \ \ \ \ \ \ \ \ \ \ \  \ \ \ \ \ \ \ \ \ \ \ \ \ \ \ \ \ \ \label{eq:MTPosterior}\\
g_{k}(z_{k}|X_{k})f_{k|k-1\!}(X_{k}|X_{k-1})\pi_{0:k-1\!}(X_{0:k-1}|z_{1:k-1})\nonumber
\end{eqnarray}
for $k\geq1$, where $X_{0:k}=(X_{0},...,X_{k})$, and $z_{1:k}=(z_{1},...,z_{k})$ is
the measurement history with $z_{k}$ denoting the measurement vector
at time $k$. Target trajectories or tracks can be accommodated in
the RFS formulation by incorporating a label in the target's state
vector \cite{Mahler2007,BTV13,Vu2014}. The multi-target
posterior (\ref{eq:MTPosterior}) then contains all information on
the random finite set of tracks, given the measurement history.
In \cite{Vu2014}, the set of tracks are estimated by simulating from
the multi-target posterior (\ref{eq:MTPosterior}) using particle Markov Chain Monte Carlo
(PMCMC) techniques \cite{Andrieu2010}. 

Computing the multi-target
posterior is prohibitively expensive for on-line applications. A more
tractable alternative is the marginal $\pi_{k}$ at time $k$ known
as the \emph{multi-target filtering density}. For notational compactness
we omit the dependence on the measurement history. Marginalizing the
multi-target posterior recursion (\ref{eq:MTPosterior}) yields the
\emph{multi-target Bayes filter }\cite{MahlerPHD2,Mahler2007},
\begin{align}
\!\!\pi_{\! k|k-1\!}(X_{k}) & =\!\int\! f_{\! k|k-1\!}(X_{k}|X)\pi_{k-1\!}(X)\delta\! X,\label{eq:MTBayesPred}\\
\!\!\pi_{k}(X_{k}) & =\!\frac{g_{k}(z_{k}|X_{k})\pi_{k|k-1}(X_{k})}{\int g_{k}(z_{k}|X)\pi_{k|k-1}(X)\delta X},\label{eq:MTBayesUpdate}
\end{align}
where $\pi_{k|k-1}$ is the \emph{multi-target prediction density}
to time $k$, and the integral is a \emph{set integral} defined for
any function $f\mathbf{:}\mathcal{F}\mathbf{(}\mathcal{X}\mathbf{)}\rightarrow\mathbb{R}$
by
\begin{equation}
\int f(X)\delta X=\sum_{i=0}^{\infty}\frac{1}{i!}\int f(\{x_{1},...,x_{i}\})d(x_{1},...,x_{i}).
\end{equation}
In \cite{BTV13,Vo2014} an analytic solution to the multi-target Bayes filter (\ref{eq:MTBayesPred}), 
(\ref{eq:MTBayesUpdate}), known as the Vo-Vo filter \cite{Mahler2014}, was derived using labeled RFSs.
Note that the majority of work in multi-target tracking
is based on filtering, and often the term \textquotedbl{}multi-target
posterior\textquotedbl{} is used in place of \textquotedbl{}multi-target
filtering density\textquotedbl{}.

\subsection{Superpositional Sensor\label{sub:Superpositional-Sensor}}

In a superpositional sensor model, the measurement $z$ is a non-linear
function of the sum of the contributions of individual targets and
noise, i.e.
\begin{equation}
z=\Phi\left(\sum\limits _{x\in X}\gamma(x),\varepsilon\right)\label{eq:SPsensor}
\end{equation}
where $\gamma(x)$ represents the contribution of the single-target
state $x$ to the sensor measurement ($\gamma$ is a non-linear mapping
in general), $\varepsilon$ is the measurement noise, and $\Phi$
is a nonlinear mapping. For example,\ a superpositional sensor model
commonly used in radar is
\begin{equation}
z=\left\vert e^{j\theta}\sum\limits _{x\in X}A(x)\varsigma(x)+\zeta\right\vert ^{2},
\end{equation}
where $\varsigma(x)$\ is the point-spread function of target $x$,
$A(x)$\ is the (known) amplitude, $\theta$ is the phase noise,
uniformly distributed on $[0,2\pi]$, and $\zeta$ is circularly complex
symmetric Gaussian noise. It is clear that this model takes on the
form (\ref{eq:SPsensor}) by defining\ $\varepsilon=(\theta,\zeta)$.
In general, the multi-target likelihood function for the superpositional
sensor model is the probability density of the measurement $z$ given $\sum\nolimits _{x\in X}\gamma(x),$ the sum of the contributions
of individual targets, i.e.
\begin{equation}
g_{k}(z|X)=h_{k}\left(z;\sum\limits _{x\in X}\gamma(x)\right)\mathbf{,}\label{eq:SPlikelihood}
\end{equation}

\noindent The SA-CPHD filter filter presented in \cite{Nan13}
is an approximation to the multi-target Bayes filter for a superpositional
measurement model of the form
\begin{equation}
z=\sum\limits _{x\in X}\gamma(x)+\varepsilon,\label{eq:SCPHDlikelihood}
\end{equation}
where\ $\varepsilon$ is distributed according to $\mathcal{N}_{R}$,
a zero mean Gaussian with covariance $R$. Hence, the likelihood
function for the superpositional measurement is
\begin{equation}
g_{k}(z|X)=\mathcal{N}_{R}\left(z-\sum\limits _{x\in X}\gamma(x)\right)\mathbf{.}
\end{equation}
Similar to the CPHD filter (for the standard sensor model) \cite{MahlerCPHD2007},
the SA-CPHD filter filter \cite{Nan13} is an analytic approximation
of the Bayes multi-target filter (\ref{eq:MTBayesPred}), (\ref{eq:MTBayesUpdate})
based on independently and identically distributed (iid) cluster RFS.
A brief review of the SA-CPHD filter is given in subsection
\ref{sub:Superpositional-Approximate-CPHD}. Both filters recursively
propagate the cardinality distribution and the PHD of the posterior
multi-target RFS. The CPHD filter can be implemented with Gaussian
mixtures or particles \cite{VVC07}, while only the
particle implementation is available for the SA-CPHD
filter \cite{Nan13}. Particle implementations of PHD/CPHD filter
in general require clustering to extract multi-target estimates, which
can introduce additional errors under challenging scenarios.

\subsection{Labeled RFS \label{sub:Labeled-RFS}}

To perform tracking in the RFS framework we use the labeled RFS model
which incorporates a unique label in the target's state vector to
identify its trajectory \cite{Mahler2007}. In this model,
the single-target state space $\mathcal{X}$ is a Cartesian product
$\mathbb{X}\mathcal{\times}\mathbb{L}$, where $\mathbb{X}$ is the
feature/kinematic space and $\mathbb{L}$ is the (discrete) label
space. A finite subset set $\mathbf{X}$ of $\mathbb{X}\mathcal{\times}\mathbb{L}$
has distinct labels if and only if $\mathbf{X}$ and its labels $\{\ell\!:\!(x,\ell)\!\in\!\mathbf{X}\}$
have the same cardinality. An RFS on $\mathbb{X}\mathcal{\times}\mathbb{L}$
with distinct labels is called a \emph{labeled RFS} \cite{BTV13,Vo2014}.

For the rest of the paper, we use the standard inner product notation
$\left\langle f,g\right\rangle \triangleq\int f(x)g(x)dx$. We denote
a generalization of the Kroneker delta and the inclusion function
that take arbitrary arguments such as sets, vectors, by
\begin{equation*}
\delta_{Y}(X)  \triangleq  \left\{ \begin{array}{l}
1,\text{ if }X=Y\\
0,\text{ otherwise}
\end{array}\right.,~~~~
1_{Y}(X)  \triangleq  \left\{ \begin{array}{l}
1,\text{ if }X\subseteq Y\\
0,\text{ otherwise}
\end{array}\right..
\end{equation*}
We also write $1_{Y}(x)$ in place of $1_{Y}(\{x\})$ when $X$ =
$\{x\}$. Single-target states are represented by lowercase letters,
e.g. $x$, $\mathbf{x}$ while multi-target states are represented
by uppercase letters, e.g. $X$, $\mathbf{X}$, symbols for labeled
states and their distributions are bolded to distinguish them from
unlabeled ones, e.g. $\mathbf{x}$, $\mathbf{X}$, $\boldsymbol{\pi}$,
etc, spaces are represented by blackboard bold e.g. $\mathbb{X}$,
$\mathbb{Z}$, $\mathbb{L}$, etc.

An important class of labeled RFS distribution is the \emph{generalized labeled
multi-Bernoulli} distribution \cite{BTV13}, known as the Vo-Vo distribution \cite{Mahler2014},
which is the basis of an analytic solution to the Bayes multi-target filter \cite{Vo2014}.
Under the standard multi-target measurement model, the Vo-Vo distribution is a conjugate
prior that is also closed under the Chapman-Kolmogorov equation. If
we start with a Vo-Vo initial prior, then the multi-target posterior
at any time is a also a Vo-Vo distribution.
Let $\mathcal{L}:\mathbb{X}\mathcal{\times}\mathbb{L}\rightarrow\mathbb{L}$
be the projection $\mathcal{L}((x,\ell))=\ell$, let $\Delta(\mathbf{X})\triangleq$
$\delta_{|\mathbf{X}|}(|\mathcal{L}(\mathbf{X})|)$ denote the \emph{distinct
label indicator}, and $h^{X}\triangleq\underset{x\in X}{\prod}h(x)$,
denote the multi-object exponential, where $h$ is a real-valued function,
with $h^{\emptyset}=1$ by convention. A Vo-Vo density is a labeled RFS density on 
$\mathcal{F}(\mathbb{X}\mathcal{\times}\mathbb{L})$ 
\begin{equation}
\boldsymbol{\pi}(\mathbf{X})=\Delta(\mathbf{X})\sum_{c\in\mathbb{C}}\omega^{(c)}(\mathcal{L}(\mathbf{X}))\left[p^{(c)}\right]^{\mathbf{X}}\label{eq:GLMB}
\end{equation}
where $\mathbb{C}$ is a discrete index set, $w^{(c)}(L)$ and $p^{(c)}$
satisfy:
\begin{eqnarray}
\sum_{L\subseteq\mathbb{L}}\sum_{c\in\mathbb{C}}\omega^{(c)}(L) & = & 1,\\
\int p^{(c)}(x,\ell)dx & = & 1.
\end{eqnarray}
The Vo-Vo density (\ref{eq:GLMB}) can be interpreted as a mixture
of multi-object exponentials. Each term in (\ref{eq:GLMB}) consists
of a weight $\omega^{(c)}(\mathcal{L}(\mathbf{X}))$ that depends
only on the labels of $\mathbf{X}$, and a multi-object exponential
$\left[p^{(c)}\right]^{\mathbf{X}}$ that depends on the entire $\mathbf{X}$.
The Labeled Multi-Bernoulli (LMB) family is a special case of the
Vo-Vo density with one term of the form:
\begin{eqnarray}
\boldsymbol{\pi}(\mathbf{X}) & = & \Delta(\mathbf{X})\omega(\mathcal{L}(\mathbf{X}))p^{\mathbf{X}}\label{eq:LMB}\\
p(x,\ell) & = & p^{(\ell)}(x)\\
\omega(L) & = & \prod\limits _{\ell\in\mathbb{M}}\left(1-r^{(\ell)}\right)\prod\limits _{\ell\in L}\frac{1_{\mathbb{M}}(\ell)r^{(\ell)}}{1-r^{(\ell)}}
\end{eqnarray}
where $\{(r^{(\ell)},p^{(\ell)})\}_{\ell\in\mathbb{M}}$, $\mathbb{M\subseteq L}$,
is a given set of parameters with $r^{(\ell)}$ representing the existence
probability of track $\ell$, and $p^{(\ell)}$ the probability density
of the kinematic state of track $\ell$ given its existence \cite{BTV13}.
Note that for an LMB the index space $\mathbb{C}$ has only one element,
in which case the $(c)$\ superscript is not needed. The LMB family
is the basis of the LMB filter, a principled and efficient approximation
of the Bayes multi-target tracking filter, which is highly parallelizable
and capable of tracking large number of targets \cite{Reuter2014}.

\section{Bayesian multi-target tracking for superpositional sensor\label{sec:Multitarget_tracker}}

In this section we describe the classical particle Bayes multi-target
filter \cite{VSD05}, which has very high computational complexity
in general. Fortunately, using labeled targets greatly simplifies
the multi-target transition density and drastically reduces the computational
complexity. Subsection \ref{subsec:RFSBayes} presents a summary of
the classical multi-target particle filter and Subsection \ref{sub:Labeled-multi-target-transition}
details the labeled multi-target transition density that reduces the
computational complexity. Subsection \ref{sub:Superpositional-Approximate-CPHD}
reviews the equations of the superpositional CPHD filter that is used
to following section to construct LMB/Vo-Vo efficient proposal distribution
for the multi-target particle filter.

Following \cite{BTV13,Vo2014}, to ensure distinct labels we assign
each target an ordered pair of integers $\ell=(k,i)$, where $k$
is the time of birth and $i$ is a unique index to distinguish targets
born at the same time. The label space for targets born at time $k$
is denoted as $\mathbb{L}_{k}$, and a target born at time $k$, has
state $\mathbf{x}\in\mathbb{X}\mathcal{\times}\mathbb{L}_{k}$. The
label space for targets at time $k$ (including those born prior to
$k$), denoted as $\mathbb{L}_{0:k}$, is constructed recursively
by $\mathbb{L}_{0:k}=\mathbb{L}_{0:k-1}\cup\mathbb{L}_{k}$ (note
that $\mathbb{L}_{0:k-1}$ and $\mathbb{L}_{k}$ are disjoint). A
multi-target state $\mathbf{X}$ at time $k$, is a finite subset
of $\mathcal{X=}$ $\mathbb{X}\mathcal{\times}\mathbb{L}_{0:k}$.
For completeness, the Bayes multi-target tracking filter, i.e. the
multi-target Bayes recursion (\ref{eq:MTBayesPred}), (\ref{eq:MTBayesUpdate})
for labeled RFS, is provided below
\begin{align}
\!\!\boldsymbol{\pi}_{\! k|k-1\!}(\mathbf{X}_{k}) & =\!\int\!\mathbf{f}_{\! k|k-1\!}(\mathbf{X}_{k}|\mathbf{X})\boldsymbol{\pi}_{k-1\!}(\mathbf{X})\delta\mathbf{X},\\
\boldsymbol{\pi}_{k}(\mathbf{X}_{k}) & =\!\frac{g_{k}(z_{k}|\mathbf{X}_{k})\boldsymbol{\pi}_{k|k-1}(\mathbf{X}_{k})}{\int g_{k}(z_{k}|\mathbf{X})\boldsymbol{\pi}_{k|k-1}(\mathbf{X})\delta\mathbf{X}}.
\end{align}

\subsection{Particle Bayes multi-target filter \label{subsec:RFSBayes}}

The propagation of the multi-target posterior involves the evaluation
of multiple set integrals and hence the computational requirement
is much more intensive than single-target filtering. Particle filtering
techniques permit recursive propagation of the set of weighted particles
that approximate the posterior. Central in Monte Carlo methods is
the notion of approximating the integrals of interest using random
samples. While the FISST density is not a density (in the Radon-Nikodym
context), it can be converted into a probability density (with respect
to a particular dominating measure) by cancelling out the unit of
measurement \cite{VSD05}. Monte Carlo approximations of the integrals
of interest can then be constructed using random samples. The single-target
particle filter can thus be directly generalised to the multi-target
case. In the multi-target context however, each particle is a finite
set and the particles themselves can thus be of varying dimensions.
Following \cite{VSD05}, suppose that at time $k-1$, a set of weighted
particles $\{w_{k-1}^{(i)},\mathbf{X}_{k-1}^{(i)}\}_{i=1}^{N_{p}}$
representing the multi-target posterior $\boldsymbol{\pi}_{k-1|k-1}$
is available, i.e.
\begin{equation}
\boldsymbol{\pi}_{k-1}(\mathbf{X})\approx\sum_{i=1}^{N_{p}}w_{k-1}^{(i)}\delta(\mathbf{X};\mathbf{X}_{k-1}^{(i)})
\end{equation}
Note that $\delta(\mathbf{\cdot;X}_{k-1}^{(i)})$ is the Dirac-delta
concentrated at $\mathbf{X}_{k-1}^{(i)}$ (different from the Kronecker-delta
$\delta_{X}$ that takes values of either 1 or 0). The particle filter
proceeds to approximate the multi-target posterior $\boldsymbol{\pi}_{k}$
at time $k$ by a new set of weighted particles $\{w_{k}^{(i)},\mathbf{X}_{k}^{(i)}\}_{i=1}^{N_{p}}$
as follows

\vspace{2mm}

\begin{center}
\textbf{Multi-target Particle Filter}
\par\end{center}

\hrule

\vspace{1mm}

\noindent \begin{flushleft}
\textsf{\uline{\footnotesize{For time $k\geq1$}}}
\par\end{flushleft}{\footnotesize \par}
\begin{itemize}
\item \textsf{\footnotesize{For $i=1,\ldots,N_{p}$}}{\footnotesize{ sample
$\tilde{\mathbf{X}}_{k}^{(i)}\sim\mathbf{q}(\cdot|\mathbf{X}_{k-1}^{(i)},z_{k})$
and set
\begin{equation}
\tilde{w}_{k}^{(i)}=\frac{g_{k}(z_{k}|\tilde{\mathbf{X}}_{k}^{(i)})\mathbf{f}_{k|k-1}(\tilde{\mathbf{X}}_{k}^{(i)}|\mathbf{X}_{k-1}^{(i)})}{\mathbf{q}_{k}(\tilde{\mathbf{X}}_{k}^{(i)}|\mathbf{X}_{k-1}^{(i)},z_{k})}w_{k-1}^{(i)}\label{eq:W_update}
\end{equation}
}}{\footnotesize \par}
\item {\footnotesize{Normalize the weights: $\tilde{w}_{k}^{(i)}=\frac{\tilde{w}_{k}^{(i)}}{\sum_{i=1}^{N_{p}}\tilde{w}_{k}^{(i)}}$ }}{\footnotesize \par}
\end{itemize}
\uline{\footnotesize{Resampling Step}}{\footnotesize \par}
\begin{itemize}
\item {\footnotesize{Resample $\{\tilde{w}_{k}^{(i)},\tilde{\mathbf{X}}_{k}^{(i)}\}_{i=1}^{N_{p}}$
to get $\{w_{k}^{(i)},\mathbf{X}_{k}^{(i)}\}_{i=1}^{N_{p}}$ }}
\end{itemize}
\vspace{1mm}\hrule\vspace{1mm}

The importance sampling density $\mathbf{q}_{k}(\cdot|\mathbf{X}_{k-1},z_{k})$
is a multi-target density and $\tilde{\mathbf{X}}_{k}$ is a sample
from an RFS. It is implicit in the above algorithm description that
\begin{equation}
\sup_{\mathbf{X}_{k},\mathbf{X}_{k-1}}\left\vert \frac{\mathbf{f}_{k|k-1}(\mathbf{X}_{k}|\mathbf{X}_{k-1})}{\mathbf{q}_{k}(\mathbf{X}_{k}|\mathbf{X}_{k-1},z_{k})}\right\vert <\infty
\end{equation}
so that the weights are well-defined. Convergence results for the
multi-target particle filter are given in \cite{VSD05}.

Notice that the entire posterior can be computed by modifying the pseudo-code 
of the multi-target particle filter so that $\mathbf{X}^{(i)}_{0:k}$ is used in place of $\mathbf{X}^{(i)}_{k}$
and $\mathbf{X}^{(i)}_{0:k-1}$ is used in place of $\mathbf{X}^{(i)}_{0:k-1}$.
This would in principle solve the so called mixed labelling problem \cite{Boers2007}. 
However, this is computationally demanding because it
requires recomputing the whole history of each multi-target particle \cite{Vu2014}. 
Alternatively, forward-backward smoothing can be used to approximate the entire posterior \cite{Mahler2007}.
In this paper we focus on designing efficient proposal distributions for the multi-target
particle filter approximating the filtering recursion. In future work we will consider the
application of the proposed approach to the problem of estimating the full posterior. 

The main practical problem with the multi-target particle filter is
the need to perform importance sampling in very high dimensional spaces
if many targets are present. In \cite{Ma2006,Ristic2010,Reuter2013},
the transition density is used as the proposal, i.e. $\mathbf{q}_{k}(\cdot|\mathbf{X}_{k-1}^{(i)},z_{k})=\mathbf{f}_{k|k-1}(\cdot|\mathbf{X}_{k-1}^{(i)})$.
While this avoids the evaluation of the transition density, it suffers
from particle depletion even for a small number of targets. This problem
is compounded with superpositional sensor due to less informative
measurements arising from low SNR. A naive choice of importance density
such as the transition density will typically lead to an algorithm
whose efficiency decreases exponentially with the number of targets
for a fixed number of particles \cite{VSD05}. The problem with using
a proposal other than the transition density, is that the weights
are difficult to evaluate due to the combinatorial nature of the transition
density for unlabled RFS. Fortunately, for labeled RFS the transition
density simplifies to a form that is inexpensive to evaluate.

\subsection{Labeled multi-target transition density\label{sub:Labeled-multi-target-transition}}

The multi-target transition model for labeled RFS is summarised as
follows. Given a multi-target state $\mathbf{X}$\ at time $k-1$,
each state $(x,\ell)$ $\in\mathbf{X}$ either continues to exist
at the next time step with probability $p_{S}(x,\ell)$ and evolves
to a new state $(x_{k},\ell_{k})$ with probability density $f(x_{k}|x,\ell)\delta_{\ell}(\ell_{k})$,
or dies\ with probability $1-p_{S}(x,\ell)$. In addition, the set
of new targets born at time $k$ is distributed according to the LMB distribution
\begin{equation}
\mathbf{b}_{k}(\mathbf{Y})=\Delta(\mathbf{Y})\omega_{B,k}(\mathcal{L}(Y))\left[p_{B,k}\right]^{\mathbf{Y}}\label{eq:Birth_transition}
\end{equation}
where $w_{B,k}$ and $p_{B,k}$\ are given parameters of the multi-target
birth density $\mathbf{b}_{k}$, defined on $\mathcal{F}\mathbf{(}\mathbb{X}\times\mathbb{L}_{k})$.
Note that $\mathbf{b}_{k}(\mathbf{Y})=0$ if $\mathbf{Y}$ contains
any element $\mathbf{y}$ with $\mathcal{L(}\mathbf{y})\notin\mathbb{L}_{k}$.
The birth model (\ref{eq:Birth_transition}) covers both labeled Poisson
and labeled multi-Bernoulli \cite{BTV13}. The multi-target state
$\mathbf{X}_{k}$, at time $k$, is the superposition of surviving
targets and new born targets. The model uses the standard assumption
that targets evolve independently of each other and that births are
independent of surviving targets.

It was shown in \cite{BTV13} that the multi-target transition density
is given by
\begin{align}\nonumber
\mathbf{f}_{k|k-1}&\left(\mathbf{X}_{k}|\mathbf{X}\right) = \\ & \mathbf{s}_{k|k-1}(\mathbf{X}_{k}\cap(\mathbb{X}\times\mathbb{L}_{0:k-1})|\mathbf{X})\mathbf{b}_{k}(\mathbf{X}_{k}\cap(\mathbb{X}\times\mathbb{L}_{k})) \label{eq:labeled_transition}
\end{align}
where
\small
\begin{align}\label{eq:Survival_transition}
    \mathbf{s}_{k|k-1}(\mathbf{W}|\mathbf{X}) & =  \Delta(\mathbf{W})\Delta(\mathbf{X})1_{\mathcal{L}(\mathbf{X})}(\mathcal{L}(\mathbf{W}))
	\left[\Phi_{k|k-1}(\mathbf{W};\cdot)\right]^{\mathbf{X}}	\\ \label{eq:Survival_transition2}
	\Phi_{k|k-1}(\mathbf{W};x,\ell) & = \left\{\begin{array}{ll} p_{S}(x,\ell)\ f(x_{k}|x,\ell), 
						& \!\!\text{if }\left(x_{k},\ell\right)\in\mathbf{W}\\
					    1-p_{S}(x,\ell), & \!\!\text{if }\ell\notin\mathcal{L}(\mathbf{W})
	  \end{array}\right.
\end{align}
\normalsize
Unlike the general multi-target transition density (see \cite{MahlerPHD2,Mahler2007}),
the special case for labeled RFS (\ref{eq:labeled_transition}) does
not contain any combinatorial sums. It is simply a product of terms
corresponding to the surviving targets and new targets. Consequently,
numerical complexity of the weight update in the multi-target particle
filter drastically reduces.

\subsection{{\normalsize{Superpositional Approximate CPHD filter\label{sub:Superpositional-Approximate-CPHD}}}}

In this section we recall the approximate CPHD for superpositional
measurements of the following form:
\begin{equation}
z_k=\left|\sum_{\mathbf{x}\in\mathbf{X}_{k}}h(\mathbf{x})\right|^{2}+\mathbf{n}_{k}\label{eq:ApproxTBD}
\end{equation}
where $\mathbf{X}_{k}$ is the multi-target state at time $k$, $\mathbf{n}_{k}\sim\mathcal{N}(0,\sigma_{n}^{2})$
is zero-mean white Gaussian noise, and $h(\mathbf{x})$ is
a possibly nonlinear function of the single state vector $\mathbf{x}$.
Notice that the model in eq. (\ref{eq:ApproxTBD}) can be used to
approximate the radar power measurement eq. (\ref{eq:TBD1})
assuming a Gaussian noise in power. Obviously the model in eq. (\ref{eq:ApproxTBD})
is a strong approximation of eq. (\ref{eq:TBD1}). However, it allows
using the update step of the SA-CPHD filter to
evaluate measurement updated intensity function $v(\mathbf{x})$ and
cardinality distribution $\rho(n)$ for the target set. In turn,
the information in the updated $v(\mathbf{x})$ and $\rho(n)$, along
with the targets labels from the previous step and birth process,
can be used to construct an approximate posterior density using the
Vo-Vo and/or LMB distributions in eq. (\ref{eq:GLMB}) and (\ref{eq:LMB}).
Finally, the obtained approximate posterior is used as a proposal
distribution for the multi-object particle filter.

The vector measurement $z_k$ in eq. (\ref{eq:ApproxTBD}) usually represents an array of 
sensors for SA-CPHD filter, e.g. acoustic amplitude sensors, 
radio-frequency tomography, etc.
For application of the SA-CPHD filter to tracking using radar
power returns, the vector measurement $z_k$ contains the radar power returns from the set 
of cells being interrogated by the radar at time $k$.
Hence, ${z}_{k}=[z_{k}^{(1)}\ldots z_{k}^{({m})}]$
where ${m}$ is the number of cells being interrogated by the radar.
Following \cite{Nan13}, standard CPHD
formulas are used for the predicted cardinality distribution and PHD,
while the update step of the SA-CPHD filter is given by:
\begin{align}
\rho_{k}(n)= & \rho_{k|k-1}(n)\frac{\mathcal{N}_{\Sigma_{r}+\Sigma_{k|k-1}^{n}}\left(z_{k}-n\hat{\mu}_{k|k-1}\right)}{\mathcal{N}_{\Sigma_{r}+\Sigma_{k|k-1}}\left(z_{k}-N_{k|k-1}\hat{\mu}_{k|k-1}\right)}\\
v_{k}(x)= & v_{k|k-1}(x)\frac{\mathcal{N}_{\Sigma_{r}+\Sigma_{k|k-1}^{o}}\left(z_{k}-h(x)-\mu_{k|k-1}^{o}\right)}{\mathcal{N}_{\Sigma_{r}+\Sigma_{k|k-1}}\left(z_{k}-N_{k|k-1}\hat{\mu}_{k|k-1}\right)}
\end{align}
where $\Sigma_{r}=\sigma_{n}I_{\tilde{m}}$ is the noise covariance, 
$N_{k|k-1}$ is the predicted average number of targets and: 
\small
\begin{align}
\hat{\mathbf{\mu}}_{k}= & \int h(x)s_{k|k-1}(x)dx\label{eq:muh}\\
\hat{\Sigma}_{k}= & \int h(x)h(x)^{\intercal}s_{k|k-1}(x)dx\\
\Sigma_{k}^{n}= & n\left(\hat{\Sigma}_{k}-\hat{\mathbf{\mu}}_{k}\hat{\mathbf{\mu}}_{k}^{\intercal}\right)\\
\Sigma_{k}= & N_{k|k-1}\hat{\Sigma}_{k}+(\sigma_{k|k-1}^{2}-N_{k|k-1}\hat{\mathbf{\mu}}_{k}\hat{\mathbf{\mu}}_{k}^{\intercal})\\
\mathbf{\mu}_{k}^{o}= & \frac{G_{k|k-1}^{(2)}(1)}{N_{k|k-1}}\hat{\mathbf{\mu}}_{k}\\
\Sigma_{k}^{o}= & \frac{G_{k|k-1}^{(2)}(1)}{N_{k|k-1}}\hat{\Sigma}_{k}+\left(\frac{G_{k|k-1}^{(3)}(1)}{N_{k|k-1}}-\left(\frac{G_{k|k-1}^{(2)}(1)}{N_{k|k-1}}\right)^{2}\right)\hat{\mathbf{\mu}}_{k}\hat{\mathbf{\mu}}_{k}^{\intercal}\label{eq:sigmao}
\end{align}
\normalsize
where $s_{k|k-1}(x)$ is the normalized predicted intensity, and
$\sigma_{k|k-1}^{2}$, $G_{k|k-1}^{(2)}(1)$ and $G_{k|k-1}^{(3)}(1)$
are the variance, second factorial moment and third factorial moment
of the predicted cardinality distribution $\rho_{k|k-1}(n)$. 
The equations of the superpositional approiximate CPHD filter can be implemented efficiently
using SMC methods. In the following section we describe how the updated
PHD and cardinality distribution from the SA-CPHD filter can
be used to design efficient proposal distributions for multi-target tracking.

\section{Efficient Proposal Distributions based on Superpositional Approximate CPHD filter}\label{sec:Proposals}

In this section we detail the CPHD-based proposal distribution and
the multi-target particle filter equations. In superpositional multi-target
filtering, the multi-target posterior generally cannot be written
as a product of independent densities because the target states are
statistically dependent through the measurement update. This means
that an effective particle approximation of the posterior distribution
is of great interest. Unfortunately, designing an effective multi-object
proposal distribution $\mathbf{q}(\mathbf{X}_{k}|\mathbf{X}_{k-1},z_{k})$
is not a simple task when using superpositional sensors. In this
section we exploit the SA-CPHD filter to construct a
relatively inexpensive LMB based proposal as well as more accurate
Vo-Vo based proposal. The basic idea is to obtain the updated PHD $v_{k}(x)$
and cardinality distribution $\rho_{k}(n)$ at time $k$ from the
SA-CPHD filter and construct a proposal distribution
$\mathbf{q}(\mathbf{X}_{k}|\mathbf{X}_{k-1},z_{k})$ that exploits
the approximate posterior information contained in both the cardinality
distribution $\rho_{k}(\cdot)$ and the state samples from $v_{k}(\cdot)$.

Assume a particle representation $\left\{ \mathbf{X}_{k-1}^{(i)},w_{k-1}^{(i)}\right\} _{i=1}^{N_{p}}$
of the posterior distribution $\boldsymbol{\pi}_{k-1}(\mathbf{X})$ is available
at time $k-1$. Then, the cardinality distribution $\rho(n)$ and
the PHD $v(x)$ of the unlabeled multi-target state at time $k-1$
are given by \cite{VSD05}:

\begin{eqnarray}
\rho_{k-1}(n) & \propto & \sum_{i:\left|\mathbf{X}_{k-1}^{(i)}\right|=n}w_{k-1}^{(i)}\\
v_{k-1}(x) & = & \sum_{i=1}^{N_{p}}\sum_{\ell\in\mathcal{L}\left(\mathbf{X}_{k-1}^{(i)}\right)}w_{k-1}^{(i)}\ \delta\left(x;x_{k-1,\ell}^{(i)}\right)
\end{eqnarray}

where $x_{k-1,\ell}^{(i)}$ denotes the kinematic part of each $\left(x_{k-1}^{(i)},\ell_{k-1}^{(i)}\right)\in\mathbf{X}_{k-1}^{(i)}$,
and $\delta(\mathbf{\cdot};x_{k-1,\ell}^{(i)})$ is the Dirac-delta
concentrated at $x_{k-1,\ell}^{(i)}$. The superpositional CPHD is
then used to obtain the update cardinality distribution $\rho_{k}(n)$
and PHD $v_{k}(x)$ using the measurement $z_{k}$ collected at
time $k$. Notice that differently from standard unlabeled CPHD filtering,
there is a natural labeling/clustering of particles due to the existing
labels at time $k-1$ and the chosen $iid$ cluster process with implicit cluster 
labels for the birth model. In fact, let $v_{k}(x)$ be the updated PHD at time $k$
\begin{equation}
v_{k}(x)=\sum_{i=1}^{N_{p}}\sum_{\ell\in\mathcal{L}\left(\mathbf{X}_{k}^{(i)}\right)}w_{k}^{(i)}\ \delta\left(x;x_{k,\ell}^{(i)}\right)
\end{equation}
\noindent Then we can rewrite the (unlabeled) PHD as a sum over all labels $\mathbb{L}_{0:k}$
of labeled PHD terms $v_{k,\ell}(x)$, i.e.
\begin{eqnarray}
v_{k}(x) & = & \sum_{\ell\in\mathbb{L}_{0:k}}v_{k,\ell}(x)
\end{eqnarray}
where
\begin{equation}
v_{k,\ell}(x)=\sum_{i=1}^{N_{p}}\sum_{\ell^{'}\in\mathcal{L}\left(\mathbf{X}_{k}^{(i)}\right)}\delta_{\ell}(\ell^{'})\ w_{k}^{(i)}\ \delta\left(x;x_{k,\ell}^{(i)}\right)
\end{equation}
is the contribution to the PHD of track $\ell$. Note that the above
is not the PHD of a labeled RFS but the PHD mass from a specific label
representing a survival or birth target. This means that at time $k$
we can extract $\left|\mathbb{L}_{0:k}\right|$ clusters of particles
from the posterior PHD. Furthermore, a continuous approximation to
each cluster can be obtained by evaluating sample mean and covariance
for a Gaussian approximation to $v_{k,\ell}(\cdot)$. Alternatively,
it is possible to use kernel density estimation (KDE), however this
will not be considered in this paper. For $\ell\in\mathbb{L}_{0:k}$,
let $\mu_{k,\ell}$ and $Q_{k,\ell}$ denote the sample mean and covariance
corresponding to the PHD cluster $v_{k,\ell}(\cdot$). Hence, we approximate
the PHD clusters as follows

\begin{equation}
v_{k,\ell}(x)=p_{k}^{+}(\ell)\ \mathcal{N}\left(x;\mu_{k,\ell},Q_{k,\ell}\right)
\end{equation}

\noindent where $p_{k}^{+}(\ell)$ is the PHD mass of the $\ell^{th}$
cluster. For the sake of explicitness, in our exposition we denote
the PHD mass of survival targets as $p_{k,S}^{+}(\ell)$ and the PHD
mass of newly born targets as $p_{k,B}^{+}(\ell)$,
\begin{align}
p_{k,S}^{+}(\ell) & =\sum_{i=1}^{N_{p}}\sum_{\ell^{'}\in\mathcal{L}\left(\mathbf{X}_{k}^{(i)}\right)}\delta_{\ell}(\ell^{'})\ w_{k}^{(i)}\text{ if }\ell\in\mathbb{L}_{0:k-1}\\
p_{k,B}^{+}(\ell) & =\sum_{i=1}^{N_{p}}\sum_{\ell^{'}\in\mathcal{L}\left(\mathbf{X}_{k}^{(i)}\right)}\delta_{\ell}(\ell^{'})\ w_{k}^{(i)}\text{ if }\ell\in\mathbb{L}_{k}
\end{align}

\noindent In practice we constrain the survival and birth probabilities
$p_{k,S}^{+}(\ell)\in[p_{S,min};p_{S,max}]$ and $p_{k,B}^{+}(\ell)\in[p_{B,min};p_{B,max}]$.
The constraint $p_{\cdot,min}$ is imposed to avoid the complete loss
of a track due to errors in the CPHD update while the constraint $p_{\cdot,max}$
is required since the PHD in each track cluster can exceed $1$.

The obtained posterior cardinality and posterior target clusters can
be used to construct a proposal distribution $\mathbf{q}(\cdot|\mathbf{X}_{k-1},z_{k})$.
In the following subsections we detail two strategies for constructing
the proposal $\mathbf{q}(\cdot|\mathbf{X}_{k-1},z_{k})$ as an LMB
density of the form (\ref{eq:LMB}) and as a Vo-Vo density of the form
(\ref{eq:GLMB}).

\subsection{{\normalsize{LMB Proposal Distributions}}}
In this subsection we describe how to construct a multi-target proposal distribution 
$\mathbf{q}\left(\mathbf{X}_{k}|\mathbf{X}_{k-1},z_{k}\right)$ for the multi-target 
particle tracker by using an LMB density, i.e.
\begin{align}\nonumber
\mathbf{q}&\left(\mathbf{X}_{k}|\mathbf{X}_{k-1},z_{k}\right) = \\ & \mathbf{q}_{S}(\mathbf{X}_{k}\cap(\mathbb{X}\times\mathbb{L}_{0:k})|\mathbf{X}_{k-1})\mathbf{q}_{B}	 
	(\mathbf{X}_{k}-(\mathbb{X}\times\mathbb{L}_{0:k}))
\end{align}
where $\mathbf{q}_{S}(\mathbf{X}_{k}\cap(\mathbb{X}\times\mathbb{L}_{0:k})|\mathbf{X}_{k-1})$
and $\mathbf{q}_{B}(\mathbf{X}_{k}-(\mathbb{X}\times\mathbb{L}_{0:k}))$
are the LMB proposals for survival targets and birth targets, respectively.
Specifically, the survival and birth proposals are constructed using
the CPHD updated birth $p_{k,B}^{+}(\ell)$ and survival $p_{k,S}^{+}(\ell)$
probabilities and the Gaussian clusters $\mathcal{N}(\cdot;\mu_{k,\ell},Q_{k,\ell})$.
For the survival proposal we have,
\small
\begin{align}\nonumber
\mathbf{q}_{S}&(\mathbf{W}_{k}|\mathbf{X}_{k-1},z_{k})  = \\ & \Delta(\mathbf{W}_{k})\Delta(\mathbf{X}_{k-1})1_{\mathcal{L}(\mathbf{X}_{k-1})}
	(\mathcal{L}(\mathbf{W}_{k}))\left[\Phi_{S}(\mathbf{W}_{k};\cdot)\right]^{\mathbf{X}_{k-1}} \\
\Phi_{S}&(\mathbf{W}_{k};x,\ell)  = \left\{ \begin{array}{ll}
	p_{k,S}^{+}(\ell)\ \mathcal{N}(x;\mu_{k,\ell},Q_{k,\ell}), & \text{if }\left(x,\ell\right)\in\mathbf{W}_{k}\\
	1-p_{k,S}^{+}(\ell), & \text{if }\ell\notin\mathcal{L}(\mathbf{W}_{k})
\end{array}\right.
\end{align}
\normalsize
while for the birth proposal we have,
\begin{align}
	\mathbf{q}_{B}(\mathbf{X}_{k}) & = \Delta(\mathbf{X}_{k})\left[\Phi_{B}(\cdot)\right]^{\mathbf{X}_{k}} \\
	\Phi_{B}(\mathbf{X}_{k}) & =
	 \left\{ \begin{array}{ll} p_{k,B}^{+}(\ell)\ \mathcal{N}(x;\mu_{k,\ell},Q_{k,\ell}),
								& \text{if }\left(x,\ell\right)\in\mathbf{X}_{k}\\
								1-p_{k,B}^{+}(\ell), & \text{if }\ell\notin\mathcal{L}(\mathbf{X}_{k})
			  \end{array}\right.
\end{align}
In summary, the multi-target proposal distribution in eq. (\ref{eq:W_update}) is 
constructed using two LMB densities for the existing and newly appeared targets, respectively.
A pseudo-code of the multi-target particle filter using the LMB proposal for sampling is given
below. Notice that we used the following definitions for grouping of labels in each particle
\begin{eqnarray}
\mathcal{L}_{S}^{(i)} & = & \mathcal{L}\left(\mathbf{X}_{k-1}^{(i)}\right)\bigcap\mathcal{L}\left(\mathbf{X}_{k}^{(i)}\right)\\
\mathcal{L}_{D}^{(i)} & = &  \mathcal{L}\left(\mathbf{X}_{k-1}^{(i)}\right)-\mathcal{L}\left(\mathbf{X}_{k}^{(i)}\right)\\
\mathcal{L}_{B}^{(i)} & = & \mathbb{L}_{k}\bigcap\mathcal{L}\left(\mathbf{X}_{k}^{(i)}\right)\\
\mathcal{L}_{NB}^{(i)}&  = & \mathbb{L}_{k}-\mathcal{L}\left(\mathbf{X}_{k}^{(i)}\right)
\end{eqnarray}

where for each particle $(i)$, $\mathcal{L}_{S}^{(i)}$ is the set
of survived labels, $\mathcal{L}_{D}^{(i)}$ is the set of death labels,
$\mathcal{L}_{B}^{(i)}$ is the set of labels for newly born targets,
and $\mathcal{L}_{NB}^{(i)}$ is the set of labels that did not generate
a new targets.

\vspace{2mm}

\begin{center}
\textbf{Multi-Target Particle Filter\\ with LMB Proposal Distribution}
\par\end{center}

\hrule

\vspace{1mm}

\noindent \textsf{\footnotesize{Initialize particles $\mathbf{X}_{0}^{(i)}\sim p_{0}(\cdot)$}}{\footnotesize \par}

\noindent \textsf{\footnotesize{For $k=1,\ldots,K$}}{\footnotesize \par}
\begin{itemize}
\item \noindent \textsf{\footnotesize{For $i=1,\ldots,N_{p}$}}{\footnotesize{ }}{\footnotesize \par}

\begin{itemize}
\item \noindent \textsf{\footnotesize{For each $\ell\in\mathcal{L}\left(\mathbf{X}_{k-1}^{(i)}\right)$}}{\footnotesize{ }}{\footnotesize \par}

\begin{itemize}
\item {\footnotesize{Generate $\alpha\sim\mathcal{U}_{[0,1]}$ }}{\footnotesize \par}
\item {\footnotesize{If $\alpha\leq p_{k,S}^{+}(\ell)$ generate $x_{k,\ell}^{(i)}\sim\mathcal{N}$$\left(\mu_{k,\ell},Q_{k,\ell}\right)$}}{\footnotesize \par}
\end{itemize}
\item \textsf{\footnotesize{For each $\ell\in\mathbb{\mathbb{L}}_{k}$}}{\footnotesize{ }}{\footnotesize \par}

\begin{itemize}
\item {\footnotesize{Generate $\alpha\sim\mathcal{U}_{[0,1]}$ }}{\footnotesize \par}
\item {\footnotesize{If $\alpha\leq p_{k,B}^{+}(\ell)$ generate $x_{k,\ell}^{(i)}\sim\mathcal{N}$$\left(\mu_{k,\ell},Q_{k,\ell}\right)$}}{\footnotesize \par}
\end{itemize}
\item \textsf{\footnotesize{Evaluate the transition kernel}}{\footnotesize \par}
\end{itemize}

{\footnotesize{
	\begin{align}\nonumber
		\mathbf{f}_{k|k-1}&\left(\mathbf{X}_{k}^{(i)}|\mathbf{X}_{k-1}^{(i)}\right) = 	
		\prod_{\ell\in\mathcal{L}_{D}^{(i)}}\left(1-p_{S}(\ell)\right)\prod_{\ell\in\mathcal{L}_{NB}^{(i)}}\left(1-p_{B}(\ell)\right)\times \\ 
		& \prod_{\ell\in\mathcal{L}_{S}^{(i)}}p_{S}(\ell)~f_{k|k-1}(x_{k,\ell}^{(i)}|x_{k-1,\ell}^{(i)})
		  \prod_{\ell\in\mathcal{L}_{B}^{(i)}}p_{B}(\ell)~p_{B}(x_{k,\ell}^{(i)})
	\end{align}
}}
{\footnotesize \par}
\begin{itemize}
\item \noindent \textsf{\footnotesize{Evaluate the proposal distribution}}{\footnotesize}
\end{itemize}
\noindent 
{\footnotesize{
\begin{eqnarray*}
\quad\quad\mathbf{q}_{k|k-1}\left(\mathbf{X}_{k}^{(i)}|\mathbf{X}_{k-1}^{(i)},z_{k}\right)=\ \ \ \ \ \ \ \ \ \ \ \ \ \ \ \ \ \ \ \ \ \ \ \ \ \ \ \ \ \ \ \ \ \ \ \ \ \ \ \ \ \ \ \ \ \ \\
\prod_{\ell\in\mathcal{L}_{D}^{(i)}}\left(1-p_{k,S}^{+}(\ell)\right)\prod_{\ell\in\mathcal{L}_{NB}^{(i)}}\left(1-p_{k,B}^{+}(\ell)\right)\times\ \ \ \ \ \ \ \ \ \ \ \ \ \ \ \\
\prod_{\ell\in\mathcal{L}_{S}^{(i)}}p_{k,S}^{+}(\ell)~\mathcal{N}\left(x_{k}^{(i)};\mu_{k,\ell},Q_{k,\ell}\right)\times\ \ \ \ \ \ \ \ \ \ \ \ \ \ \ \\
\prod_{\ell\in\mathcal{L}_{B}^{(i)}}p_{k,B}^{+}(\ell)\ \mathcal{N}\left(x_{k}^{(i)};\mu_{k,\ell},Q_{k,\ell}\right)\ \ \ \ \ \ \ \ \ \ \ \ \ \ \
\end{eqnarray*}
}}

{\footnotesize \par}
\begin{itemize}
\item \noindent \textsf{\footnotesize{Evaluate the multi-object likelihood
$g_{k}\left(z_{k}|\mathbf{X}_{k}^{(i)}\right)$ }}{\footnotesize \par}
\item \noindent \textsf{\footnotesize{Update the particle weight $w_{k}^{(i)}$
using eq. (\ref{eq:W_update})}}{\footnotesize \par}
\end{itemize}
\item \noindent {\small{Normalize the weights and resample as usual}}{\small \par}
\end{itemize}
{\footnotesize{\vspace{1mm}}}{\footnotesize \par}

{\footnotesize{\hrule\vspace{1mm}}}{\footnotesize \par}

\subsection{{\normalsize{Vo-Vo Proposal Distributions}}}

The LMB proposal distribution leads to an efficient sampling strategy for the multi-target
particle filter. However, the LMB proposal does not exploit all the information from the 
SA-CPHD filter since the cardinality distribution of the LMB proposal does 
not match the cardinality distribution $\rho_{k}(n)$ from the CPHD prediction/update.
Matching of the cardinality distribution $\rho_{k}(n)$ is important if we are interested 
in designing a proposal distribution that is efficient in low SNR scenarios. 
Generally, for low SNR the Multi-Bernoulli cardinality distribution is not sufficiently
informative, so that being able to estimate a more general cardinality distribution becomes
fundamental. This reasoning is true also in classical multi-target tracking with detection
measurements, e.g. the CPHD filter outperforms the Multi-Bernoulli filter in low SNR \cite{Mahler2007}.
Hence, we seek a proposal distribution that matches the CPHD cardinality exactly
while exploiting the weights of individual labeled target clusters as computed 
from the approximate posterior PHD.
A single component Vo-Vo density can be used for this purpose,
\begin{equation}
\mathbf{q}_{k}(\mathbf{X}_{k}|\mathbf{X}_{k-1},z_{k})=\Delta(\mathbf{X}_{k})\omega(\mathcal{L}(\mathbf{X}_{k}))\left[p(\cdot)\right]^{\mathbf{X}_{k}}\label{eq:GLMBproposal}
\end{equation}

We now specify the component weight $\omega(\mathcal{L}(\mathbf{X}_{k}))$
and the multi-object exponential $\left[p(\cdot)\right]^{\mathbf{X}_{k}}$, needed to match 
the CPHD updated cardinality distribution and to account for the weights of individual target clusters.
Clearly, the single-target densities $p(\cdot)$ are obtained straightforwardly
from Gaussian clusters, i.e. 
\begin{equation}
	p_{k}(x,\ell)=\mathcal{N}(x;\mu_{k,\ell},Q_{k,\ell}),\ \ell\in\mathbb{L}_{0:k}
\end{equation}

The weight $\omega(\mathcal{L}(\mathbf{X}_{k}))$ is then chosen
to preserve the CPHD cardinality distribution, and for a
given cardinality, to sample labels proportionally to the product of the posterior PHD masses of any possible label combinations. Specifically, from the posterior
PHD mass of each cluster $p_k^+(\ell)$ we construct approximate ``existence''
probabilities as
\begin{equation}
	r_k^+(j) = \frac{p_k^+(j)}{\sum_{\ell=1}^{|\mathbb L_{0:k}|} p_k^+(\ell)}, ~~ j=1,\ldots,|\mathbb L_{0:k}|
	\label{eq:PHDdist}
\end{equation}
The cardinality of the set of labels $\mathbb{L}_{0:k}$, including
birth and survival labels, grows exponentially in time. Moreover,
in any practical implementation the use of a finite sample approximation
coupled with resampling strategies typically leads to a much smaller
unique labels set at each time $k$. Thus, eq. (\ref{eq:PHDdist}) is
implemented by considering only labels from resampled particles at time $k-1$,
\begin{align}
	r_k^+(j) = & \frac{p_k^+(j)}{\sum_{\ell=1}^{|\tilde{\mathbb L}_{0:k}|} p_k^+(\ell)}, ~~ j=1,\ldots,|\tilde{\mathbb L}_{0:k}| \\
	\tilde{\mathbb{L}}_{0:k} = & {\displaystyle \bigcup_{i=1}^{N_{p}}}\mathcal{L}(\mathbf{X}_{k-1}^{(i)})\ \bigcup\mathbb{L}_{k}
\end{align}
The weight $\omega(\mathcal{L}(\mathbf{X}_{k}))$ is then defined as
\begin{equation}
	\omega(\mathcal{L}(\mathbf{X}_{k})) = \rho_{k}(|\mathcal{L}(\mathbf{X}_{k})|)
			\frac{r_k^+(\cdot)^{\mathcal{L}(\mathbf{X}_{k})}}
				 {e_{|\mathcal{L}(\mathbf{X}_{k})|} (R_k)}
%
	\label{eq:GLMBw}
\end{equation}
where $R_k=\{r_k^+(j)\}_{j\in\tilde{\mathbb L}_{0:k}}$ denotes the set of ``existence'' probabilities for
all current tracks and $e_n(\cdot)$ is the elementary symmetric function of order $n$.
%
%
%
%
%
The construction of the proposal in (\ref{eq:GLMBproposal}) leads
a simple and efficient strategy for sampling. Specifically, to sample
from (\ref{eq:GLMBproposal}) we,
\begin{itemize}
	\item sample the cardinality $\left|\mathbf{X}_{k}^{(i)}\right|$
of the newly proposed particle according to the distribution $\rho_{k}(n)$,
	\item sample $\left|\mathbf{X}_{k}^{(i)}\right|$ labels $\mathcal{L}(\mathbf{X}_{k}^{(i)})$
from $\tilde{\mathbb{L}}_{0:k}$ using the distribution defined by $\frac{r_k^+(\cdot)^{\mathcal{L}(\mathbf{X}_{k})}}
{e_{|\mathcal{L}(\mathbf{X}_{k})|} (R_k)}$,
	\item for each $\ell\in\mathcal{L}(\mathbf{X}_{k}^{(i)})$ we sample the kinematic part 
$x_{k,\ell}^{(i)}$ from $p_{k}(\cdot,\ell)=\mathcal{N}(\cdot;\mu_{k,\ell},Q_{k,\ell})$.
\end{itemize}
A detailed pseudo-code for implementation is reported below.

\vspace{2mm}

\begin{center}
\textbf{Multi-Target Particle Filter\\ with Vo-Vo Proposal Distribution}
\par\end{center}

{\footnotesize{\hrule\vspace{1mm}}}{\footnotesize \par}

\noindent \textsf{\footnotesize{Initialize particles $\mathbf{X}_{0}^{(i)}\sim p_{0}(\cdot)$}}{\footnotesize \par}

\noindent \textsf{\footnotesize{For $k=1,\ldots,K$}}{\footnotesize \par}
\begin{itemize}
\item \noindent \textsf{\footnotesize{For $i=1,\ldots,N_{p}$}}{\footnotesize{ }}{\footnotesize \par}

\begin{itemize}
\item \textsf{\footnotesize{Sample the cardinality for the new particle}}{\footnotesize{
$\left|\mathbf{X}_{k}^{(i)}\right|\sim\rho_{k}(n)$ }}{\footnotesize \par}
\item \textsf{\footnotesize{Sample the set of labels }}{\footnotesize{$\mathcal{L}\left(\mathbf{X}_{k}^{(i)}\right)$
uniformly from }}$\tilde{\mathbb{L}}_{0:k}$
\item \textsf{\footnotesize{For each $\ell\in\mathcal{L}\left(\mathbf{X}_{k}^{(i)}\right)$}}{\footnotesize{
generate $x_{k,\ell}^{(i)}\sim\mathcal{N}\left(\cdot;\mu_{k,\ell},Q_{k,\ell}\right)$ }}{\footnotesize \par}
\item \textsf{\footnotesize{For $j=1,\ldots,N_{p}$ evaluate the transition
kernel }}{\footnotesize \par}
\end{itemize}

{\footnotesize{
\begin{eqnarray*}
\quad\quad\mathbf{f}_{k|k-1}\left(\mathbf{X}_{k}^{(i)}|\mathbf{X}_{k-1}^{(j)}\right)=\ \ \ \ \ \ \ \ \ \ \ \ \ \ \ \ \ \ \ \ \ \ \ \ \ \ \ \ \ \ \ \ \ \ \ \ \ \ \ \ \ \ \ \ \ \ \\
\prod_{\ell\in\mathcal{L}_{D}^{(i)}}\left(1-p_{S}(\ell)\right)\prod_{\ell\in\mathcal{L}_{NB}^{(i)}}\left(1-p_{B}(\ell)\right)\times\ \ \ \ \ \ \ \ \ \ \ \ \ \ \ \ \ \ \ \ \ \ \\
\prod_{\ell\in\mathcal{L}_{S}^{(i)}}p_{S}(\ell)~f_{k|k-1}(x_{k,\ell}^{(i)}|x_{k-1,\ell}^{(j)})\times\ \ \ \ \ \ \ \ \ \ \ \ \ \ \ \ \ \ \ \ \ \ \\
\prod_{\ell\in\mathcal{L}_{B}^{(i)}}p_{B}(\ell)~p_{B}(x_{k,\ell}^{(i)})\ \ \ \ \ \ \ \ \ \ \ \ \ \ \ \ \ \ \ \ \ \
\end{eqnarray*}
}}
{\footnotesize \par}
\begin{itemize}
\item \noindent \textsf{\footnotesize{Evaluate the proposal distribution}}\
{\footnotesize{
\begin{eqnarray*}
\mathbf{q}_{k|k-1}\left(\mathbf{X}_{k}^{(i)}|\mathbf{X}_{k-1},z_{k}\right)=\ \ \ \ \ \ \ \ \ \ \ \ \ \ \ \ \ \ \ \ \ \ \ \ \ \ \ \ \ \ \\
\omega\left(\mathcal{L}(\mathbf{X}_{k}^{(i)})\right)\prod_{\ell\in I}\mathcal{N}\left(x_{k}^{(i)};\mu_{k,\ell},Q_{k,\ell}\right)
\end{eqnarray*}
}}
{\footnotesize \par}
\item \noindent \textsf{\footnotesize{Evaluate the multi-object likelihood
$g_{k}\left(z_{k}|\mathbf{X}_{k}^{(i)}\right)$ }}{\footnotesize \par}
\item \noindent \textsf{\footnotesize{Update the particle weight $w_{k}^{(i)}$
using}}{\footnotesize \par}
\end{itemize}
\end{itemize}
{\footnotesize{
\[
w_{k}^{(i)}=\frac{g_{k}(z_{k}|\mathbf{X}_{k}^{(i)}){\displaystyle \sum_{j=1}^{N_{p}}\mathbf{f}_{k|k-1}(\mathbf{X}_{k}^{(i)}|\mathbf{X}_{k-1}^{(j)})w_{k-1}^{(j)}}}{\mathbf{q}_{k}(\mathbf{X}_{k}^{(i)}|\mathbf{X}_{k-1},z_{k})}
\]
}}{\footnotesize \par}
\begin{itemize}
\item \noindent {\small{Normalize the weights and resample as usual}}{\small \par}
\end{itemize}
{\footnotesize{\vspace{1mm}}}{\footnotesize \par}

{\footnotesize{\hrule\vspace{2mm}}}{\footnotesize \par}

Notice from the update step in the pseudo-code that for each particle $(i)$ we require the evaluation 
of the multi-target transition kernel with respect to the previous set of particles
\begin{align}\label{eq:sum_kernel}
	w_{k}^{(i)} \propto \sum_{j=1}^{N_{p}}\mathbf{f}_{k|k-1}(\mathbf{X}_{k}^{(i)}|\mathbf{X}_{k-1}^{(j)}).
\end{align}
This is known as sum kernel problem \cite{Gra03,Bri06,Papi2014} and is due to the fact that the our
proposal $\mathbf{q}_{k}(\mathbf{X}_{k}^{(i)}|\mathbf{X}_{k-1},z_{k})$ does not depend on the previous 
particle $\mathbf{X}_{k-1}^{(i)}$ (as in the LMB proposal) but on the whole set of particles at $k-1$.
Efficient approximation techniques exist for mitigating the computational load due to the sum kernel 
problem in (\ref{eq:sum_kernel}), see \cite{Gra03,Bri06,Papi2014}. Furthermore, a simple approximate 
solution can be obtained by using a single particle $\mathbf{X}_{k-1}^{(m)}$ to evaluate the transition kernel 
\begin{align}\label{eq:1_kernel}
	\sum_{j=1}^{N_{p}}\mathbf{f}_{k|k-1}(\mathbf{X}_{k}^{(i)}|\mathbf{X}_{k-1}^{(j)})\approx
	\mathbf{f}_{k|k-1}(\mathbf{X}_{k}^{(i)}|\mathbf{X}_{k-1}^{(m)}).
\end{align} 
However, in order to use the approximation in (\ref{eq:1_kernel}) we have to choose the index $(m)$
of the previous particle in a way that guarantees $\mathbf{f}_{k|k-1}(\mathbf{X}_{k}^{(i)}|\mathbf{X}_{k-1}^{(m)})\not=0$.
This implies that the particle $\mathbf{X}_{k-1}^{(m)}$ has to verify the condition 
$\mathcal{L}(\mathbf{X}_{k}^{(i)})\cap\mathbb{L}_{0:k-1}\subseteq\mathcal{L}(\mathbf{X}_{k-1}^{(m)})$, 
i.e. the labels set of the particle $\mathbf{X}_{k-1}^{(m)}$ includes the labels set of 
surviving targets in the sampled particle $\mathbf{X}_{k}^{(i)}$.

\section{Numerical Example }\label{sec:Simulations}
In this section we demonstrate the RFS multi-target tracker with Vo-Vo proposal
via a radar tracking application for closely spaced targets and low signal-to-noise ratio.

\subsection{Dynamic Model}
The kinematic part of the single-target labeled state vector $\mathbf{x}_{k}=(x_{k},\ell_{k})$
at time $k$ is described by $x_{k}=[\tilde{x}_{k}^{T},\zeta_{k}]^{T}$,
which comprises the planar position and velocity vectors $\tilde{x}_{k}=[p_{x,k},\dot{p}_{x,k},p_{y,k},\dot{p}_{y,k}]^{T}$
in 2D Cartesian coordinates, respectively, and the unknown modulus
of the target complex amplitude $\zeta_{k}\in\mathbb{R}$. A Nearly
Constant Velocity (NCV) model is used to describe the target dynamics,
while a zero-mean Gaussian random walk is used to model the fluctuations
in time of the target complex amplitude, i.e.,
\begin{equation}
x_{k+1}=Fx_{k}+v_{k},~~v_{k}\sim\mathcal{N}\left(0;Q\right)
\end{equation}
where the matrices $F$ and $Q$ are defined as in \cite{Pap13}.

\subsection{Measurement likelihood function}

We now describe a multi-target observation model for multi-target tracking
using radar measurements. 
A target $\mathbf{x}\in\mathbf{X}$ illuminates a set of cells $C(\mathbf{x})$, 
where $C(\mathbf{x})$ is usually referred to as the \textit{target template}.
A radar positioned at the Cartesian origin collects a vector measurement
$z=[z^{(1)}\ldots z^{(m)}]$ consisting of the power signal returns, i.e.
\begin{equation}
z^{(i)}=\left|z_{A}^{(i)}\right|^{2} = 
	    \left|\sum_{\mathbf{x\in X}:i\in C(\mathbf{x})}A(\mathbf{x})h_{A}^{(i)}(\mathbf{x})+w^{(i)}\right|^{2}\label{eq:TBD1}
\end{equation}

\noindent where $z_{A}^{(i)}$ is the complex signal in cell $(i)$, and
\begin{itemize}
\item $w^{(i)}$ is a zero-mean white circularly symmetric complex Gaussian noise
with variance $2\sigma_{w}^{2}$
\item $h_{A}^{(i)}(\mathbf{x})$ is the point spread function in
cell $(i)$ from a target with state $\mathbf{x}$
\footnotesize
\begin{align}
h_{A}^{(i)}(\mathbf{x})= 
		 \exp\left(-{\displaystyle \frac{(r_{i}-r(\mathbf{x}))^{2}}{2R}-
		{\displaystyle \frac{(d_{i}-d(\mathbf{x}))^{2}}{2D}-
		{\displaystyle \frac{(b_{i}-	b(\mathbf{x}))^{2}}{2B}}}}\right)
\end{align}
\normalsize
where $R$, $D$ and $B$ are constants related to the radar cell resolution;
$r(\mathbf{x})$, $d(\mathbf{x})$, and $b(\mathbf{x})$ are the target coordinates 
in the measurement space; 
and $r_{i},d_{i},b_{i}$ are the cell centroids. 
\item $A(\mathbf{x})$ is the complex echo of target $\mathbf{x}$, i.e.
$
A(\mathbf{x})=\bar{A}_{\mathbf{x}}e^{j\theta}+a(\mathbf{x})
$
with $\bar{A}_{\mathbf{x}}$ a known amplitude, $\theta\sim [0,2\pi)$ an unknown
phase, and $a(\mathbf{x})$ a zero-mean complex Gaussian variable with 
variance $\sigma_{a(\mathbf{x})}^{2}$.
\end{itemize}
For a non-fluctuating target amplitude (Swerling 0), $A(\mathbf{x})$ is modeled as: 
\begin{equation}
A(\mathbf{x})=\bar{A}_{\mathbf{x}}\ e^{j\theta},\ \ \theta\sim\mathcal{U}_{\lbrack0,2\pi)}
\end{equation}

\noindent Let $\hat{z}^{(i)}$ denote the deterministic part of the signal in cell
$i$:
\begin{eqnarray*}
\hat{z}^{(i)}  =  \left|\hat{z}_{A}^{(i)}\right|^{2} = 
				  \left|\sum_{\mathbf{x}\in\mathbf{X}:i\in C(\mathbf{x})}
				  \bar{A}_{\mathbf{x}}\ h_{A}^{(i)}(\mathbf{x})\right|^{2}
\end{eqnarray*}
The power measurement in cell $(i)$ can be written as: {\small{
\begin{align}
z^{(i)} & =|\hat{z}_{A}^{(i)}\ e^{j\theta}+w|^{2}\notag\\
 & =\left(\hat{z}_{A}^{(i)}\cos{(\theta)}+\Re(w)\right)^{2}+\left(\hat{z_{A}}^{(i)}\sin{(\theta)}+\Im(w)\right)^{2}\notag\\
 & =U_{R}^{2}+U_{I}^{2}\notag
\end{align}
}}{\small \par}

\noindent where $U_{R}\sim\mathcal{N}(\hat{z}_{A}^{(i)}\cos(\theta),\sigma_{w}^{2})$
and $U_{I}\sim\mathcal{N}(\hat{z}_{A}^{(i)}\sin(\theta),\sigma_{w}^{2})$
are statistically independent normal random variables. Then $\sqrt{U_{R}^{2}+U_{I}^{2}}$
has a Ricean distribution, and reduces to a Rayleigh distribution
when $\hat{z}_{A}^{(i)}=0$. 
Let $SNR$ be the signal-to-noise ratio defined in dB as
\begin{equation}
	SNR = 10~\log{\left(\frac{\bar A_\mathbf{x}^2}{2~\sigma_w^2}\right)}
\end{equation}
We can choose $\sigma_w^2=1$ so that $\bar A_\mathbf{x}=\sqrt{2~10^{SNR/10}}$.
In turn, since $\sqrt{U_{R}^{2}+U_{I}^{2}}\sim\mbox{Rice}(\hat{z}_{A}^{(i)},1)$, the measurement in each cell $z^{(i)}$
is described by a non-central chi-squared distribution with $2$ degrees of freedom and non-centrality parameter $\hat{z}_{A}^{(i)}$,
and simplifies to an exponential distribution in the case $\hat{z}_{A}^{(i)}=0$.
Then, the likelihood ratio for cell $(i)$ is given by:
\begin{equation}
	\varphi(z^{(i)}|\mathbf{X}) = \exp{\left(-0.5\hat{z}^{(i)}\right)}I_{0}\left(\sqrt{z^{(i)}\hat{z}^{(i)}}\right)
\end{equation}
where $I_{0}(\cdot)$ is the modified Bessel function.
Hence, the likelihood function for the vector measurement $z$ takes the form
\begin{equation}
	g(z|\mathbf{X}) \propto \prod_{i\in C(\mathbf X)} \varphi(z^{(i)}|\mathbf{X})
\end{equation}
where $C(\mathbf X) = C(\mathbf{x}_1)\cup C(\mathbf{x}_2)\cup\ldots\cup C(\mathbf{x}_{|\mathbf{X}|})$ is the union 
of all single-target templates, i.e. the set of measurement bins used for the measurement update.
%
%

\subsection{Simulation Results}
\begin{figure}
\begin{center}
\includegraphics[scale=.55]{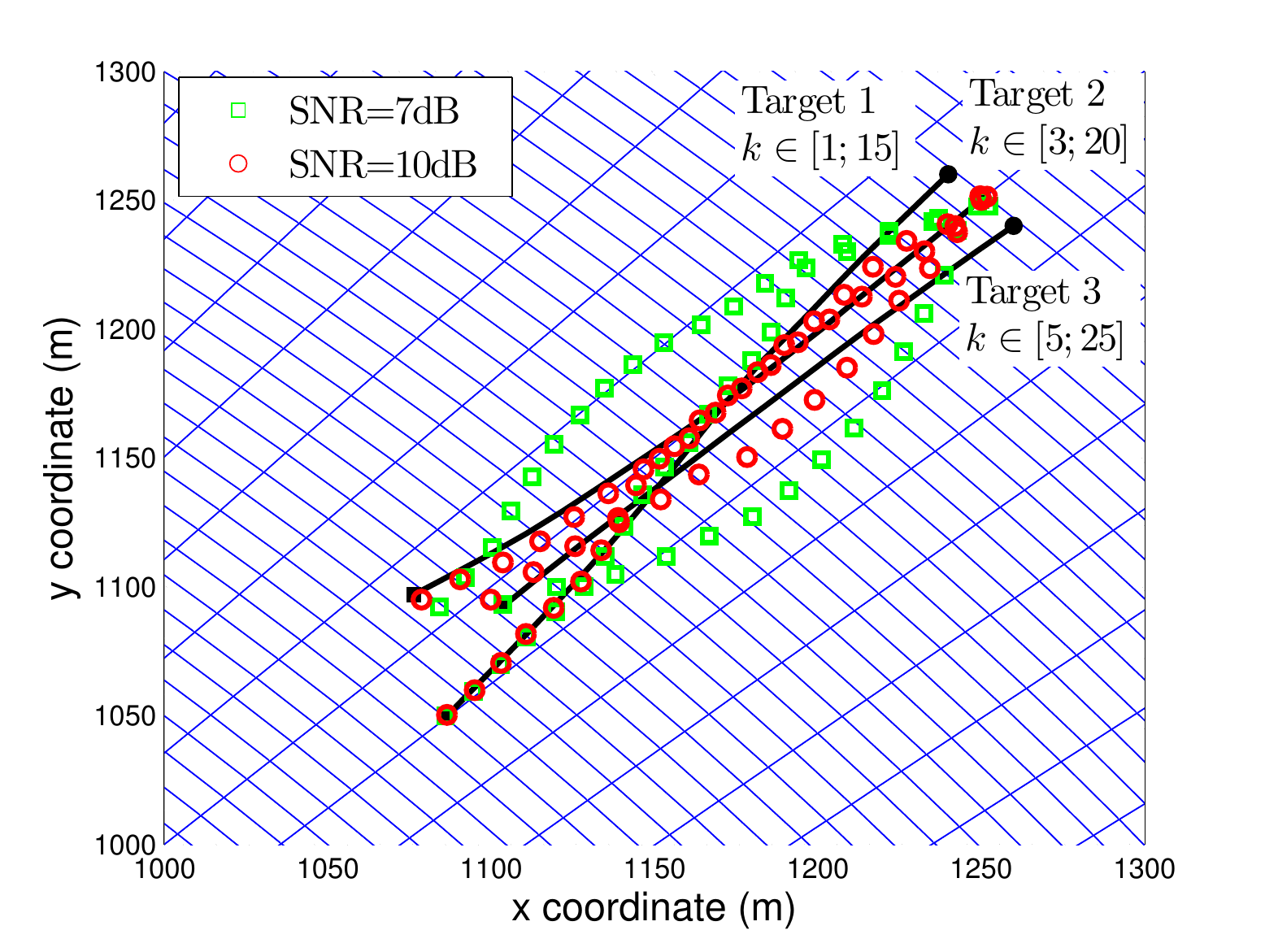}
\caption{Simulated scenario and estimated trajectories. The radar is positioned at the Cartesian origin and 
		 there are $3$ closely spaced targets moving towards the origin.}
\label{fig:scene}
\end{center}
\end{figure}

\begin{figure*}
\centering	
\subfigure[$7dB$ for $k=8$: 
		   Targets at $(r=1.7km,b=0.78\deg)$]{
	\includegraphics[width=.4\linewidth]{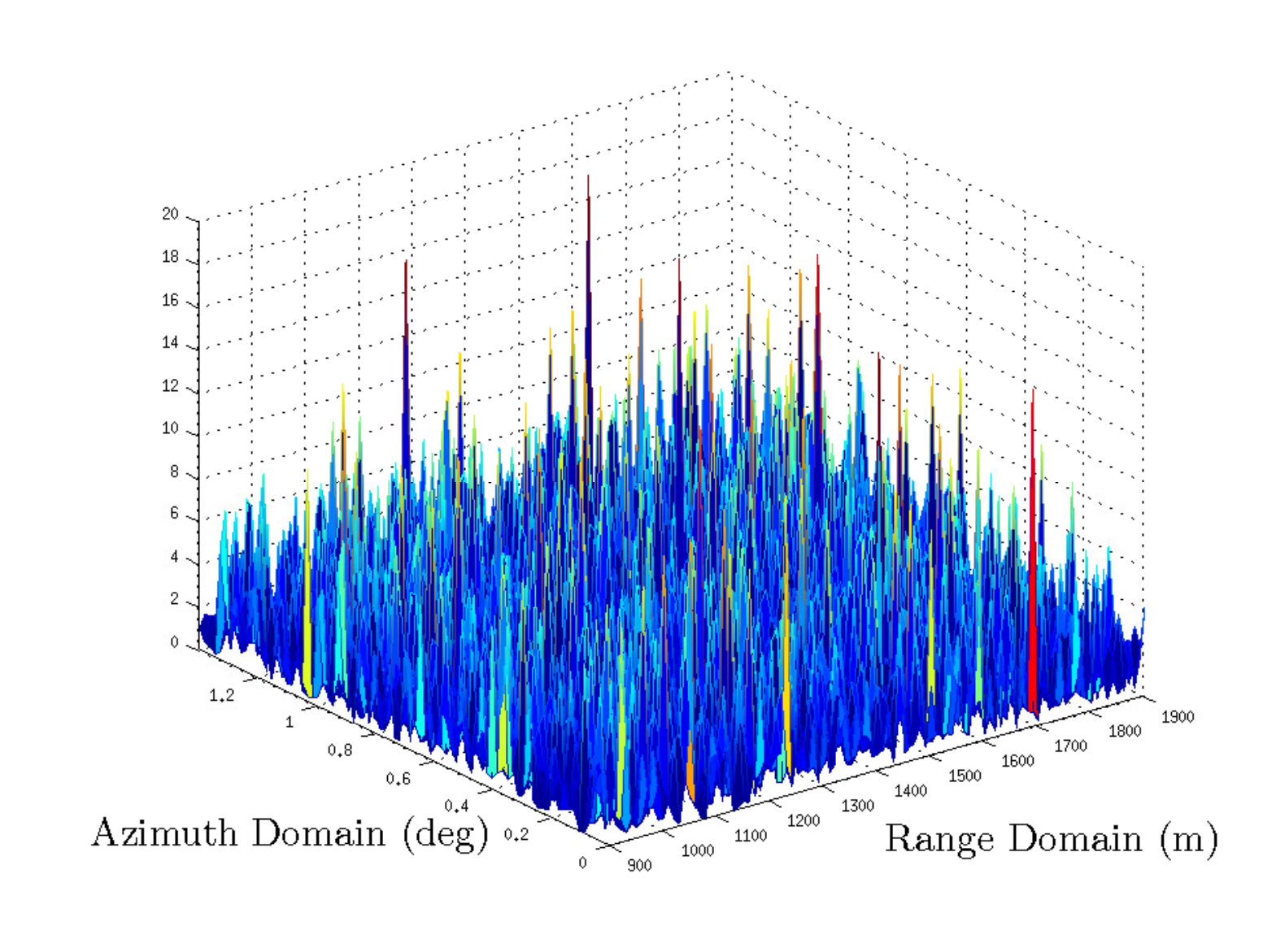}
\label{fig:meas_7dB}	
}
\hspace{2mm}
\subfigure[$10dB$ for $k=8$:
		   Targets at $(r=1.7km,b=0.78\deg)$]{
	\includegraphics[width=.4\linewidth]{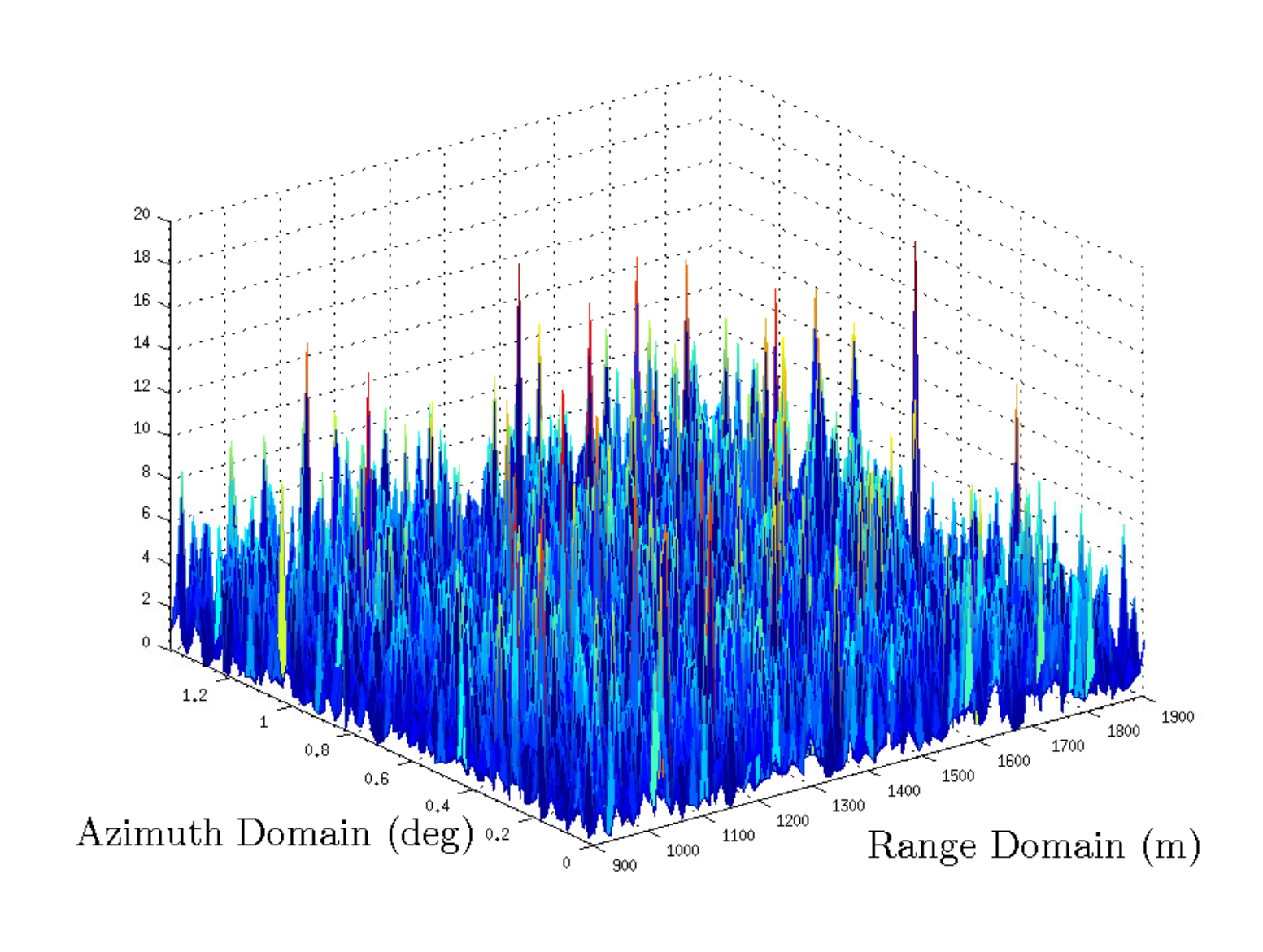}
\label{fig:meas_10dB}	
}	
\centering
\caption{Snapshot of the simulated radar power returns for low $SNR$ and closely spaced targets.}
\end{figure*}

We consider a scenario with $3$ incoming targets as depicted in Fig. \ref{fig:scene}.
To better highlight the spacing of targets, we report in Fig. \ref{fig:scene} the range-azimuth grid.
Notice how the targets share cross-range cells for most of the simulation. Consequently, a correct
estimation of the number of targets is very challenging.
In Figs. \ref{fig:meas_7dB} and \ref{fig:meas_10dB}, we report a snapshot of the linear domain power 
measurement in range-azimuth at time step $k=8$ for both cases of $SNR=7dB$ and $SNR=10dB$, respectively. 
Relevant parameters used in simulation are reported in Table \ref{tab:par}. For the filter initialization,
we describe prior knowledge using a Gaussian distribution $\mathcal{N}(\mathbf{x}_{B},Q_B)$ which mainly 
proposes incoming particles (i.e., most of particle velocity vectors are directed towards the radar position). 
Specifically, for the birth intensity we use the Gaussian $\mathcal{N}(\mathbf{x}_{B},Q_B)$
and generate $5000$ new born single-target particles in the CPHD filter at every time step. 

\begin{table}[htbp]
\caption{Common Parameters}

\footnotesize{
\begin{center}
\begin{tabular}{|c|c|c|}
\hline
Parameter & Symbol & Value\tabularnewline
\hline
\hline
Range Resolution & $R$ & $10\ m$\tabularnewline
\hline
Azimuth Resolution & $B$ & $1^{\circ}$\tabularnewline
\hline
Doppler Resolution & $D$ & $1\ m/s$\tabularnewline
\hline
Signal-to-Noise Ratio & $SNR$ & $\{7,10\} dB$\tabularnewline
\hline
Target Maximum Acceleration & $a_{max}$ & $1\ m/s$\tabularnewline
\hline
$1^{st}$ Target Initial State & $\mathbf{x}_0^1$ & $\left[1260,-11,1240,-9\right]$\tabularnewline
\hline
$2^{nd}$ Target Initial State & $\mathbf{x}_0^2$ & $\left[1250,-10,1250,-10\right]$\tabularnewline
\hline
$3^{rd}$ Target Initial State & $\mathbf{x}_0^3$ & $\left[1240,-9,1260,-11\right]$\tabularnewline
\hline
Target Birth Mean & $\mathbf{x}_{B}$ & $\left[1250,-5,1250,-5\right]$\tabularnewline
\hline
Target Birth Covariance & $Q_B$ & $\mbox{diag}\left(\left[7.5,10,7.5,10\right]^2\right)$\tabularnewline
\hline
Birth Probability & $P_{B}$ & $0.05$\tabularnewline
\hline
Survival Probability & $P_{S}$ & $0.95$\tabularnewline
\hline
$n^{\circ}$ of multi-target particles & $N_{p}$ & $\left\{3e3,5e3,10e3,20e3\right\} $\tabularnewline
\hline
\end{tabular}
\end{center}
\label{tab:par}
}
\end{table}

\begin{figure*}
\centering	
\subfigure[Monte Carlo results: No. of Targets for $SNR=10dB$]{
	\includegraphics[width=.4\linewidth]{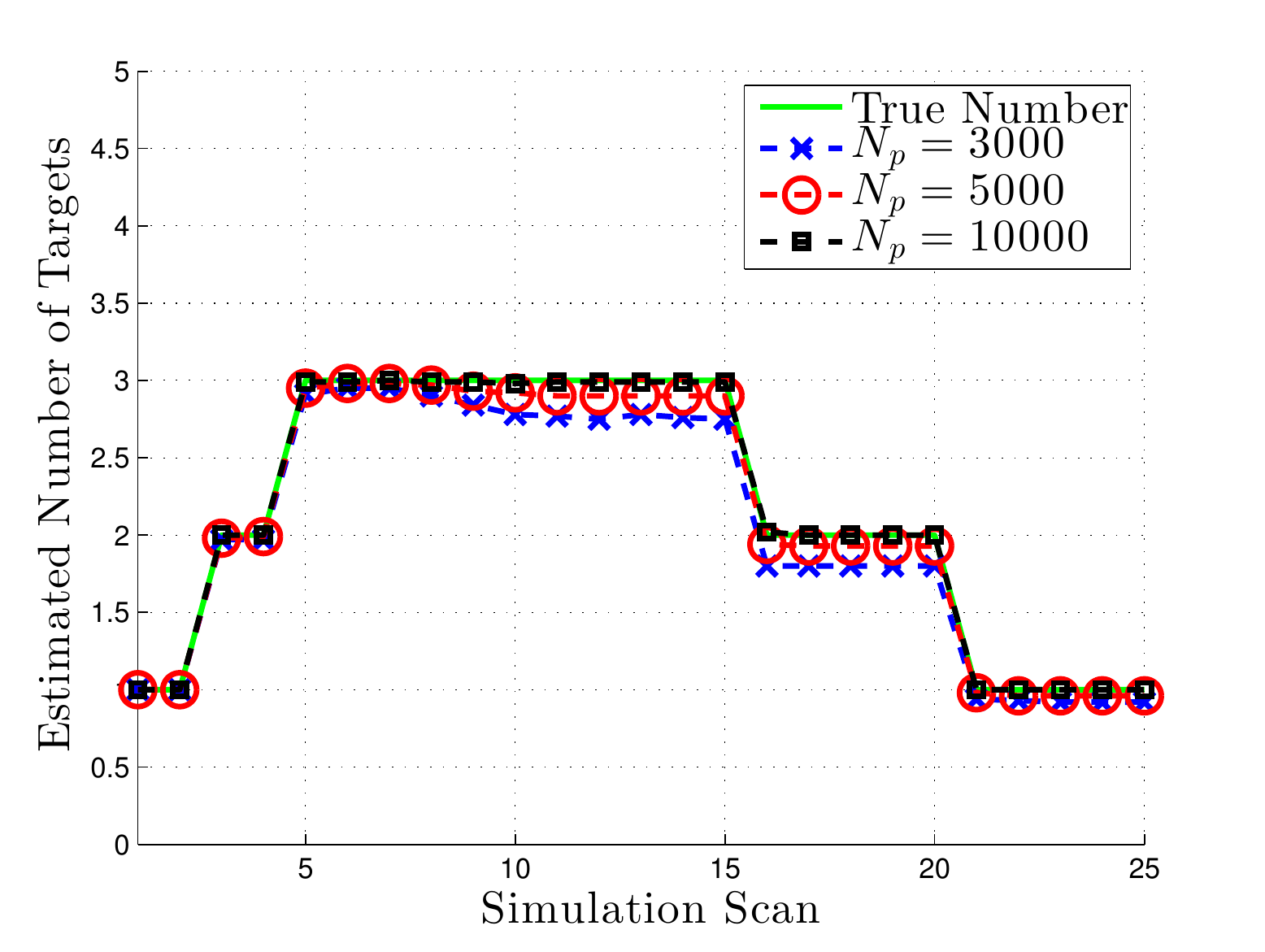}
	\label{fig:nt_10dB}
}
\hspace{2mm}
\subfigure[Monte Carlo results: OSPA Distance for $SNR=10dB$]{
	\includegraphics[width=.4\linewidth]{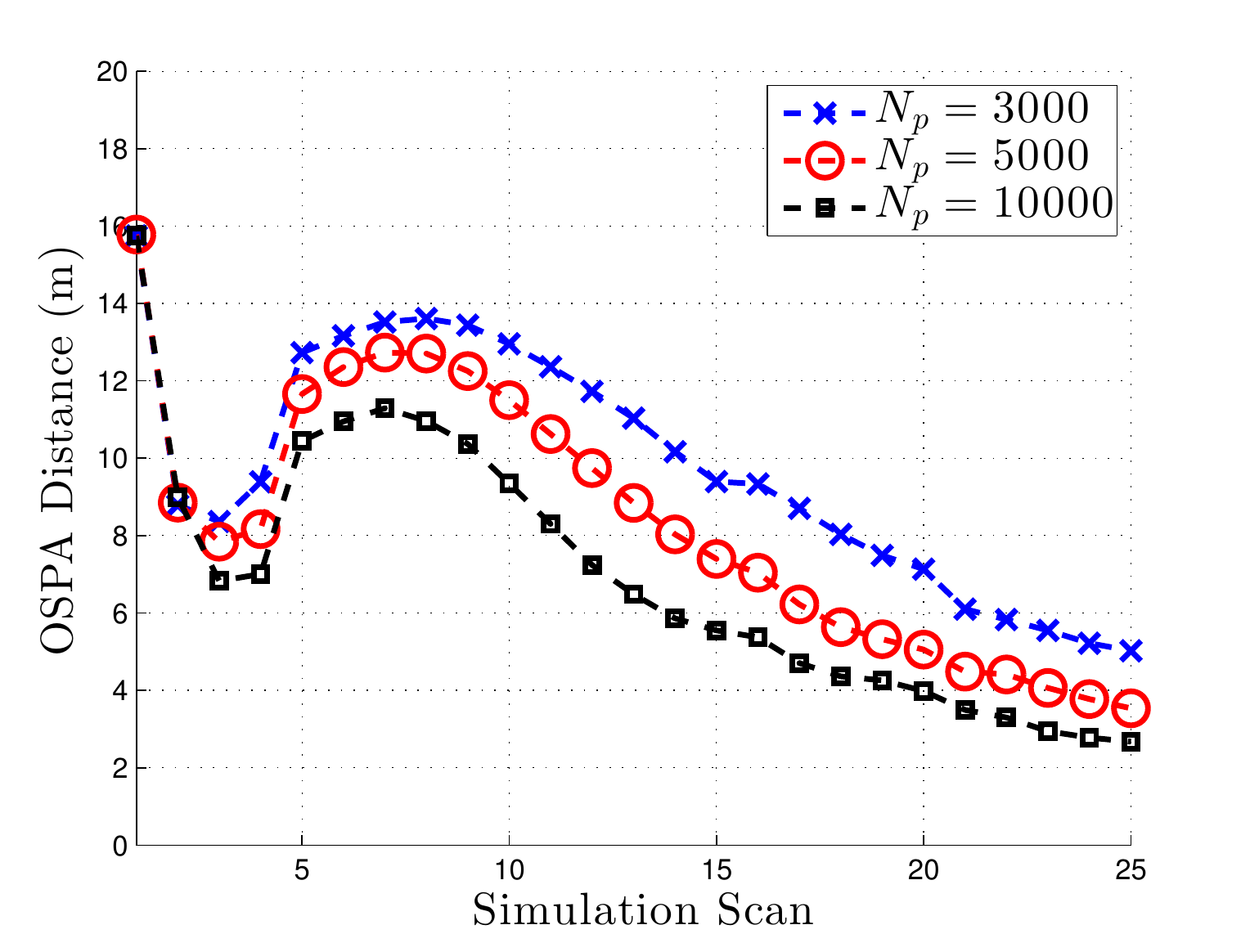}
	\label{fig:ospa_10dB}
}
\centering
\caption{Monte Carlo results for the case $SNR=10dB$}
\end{figure*}

\begin{figure*}
\centering	
\subfigure[Monte Carlo results: No. of Targets for $SNR=7dB$]{
	\includegraphics[width=.4\linewidth]{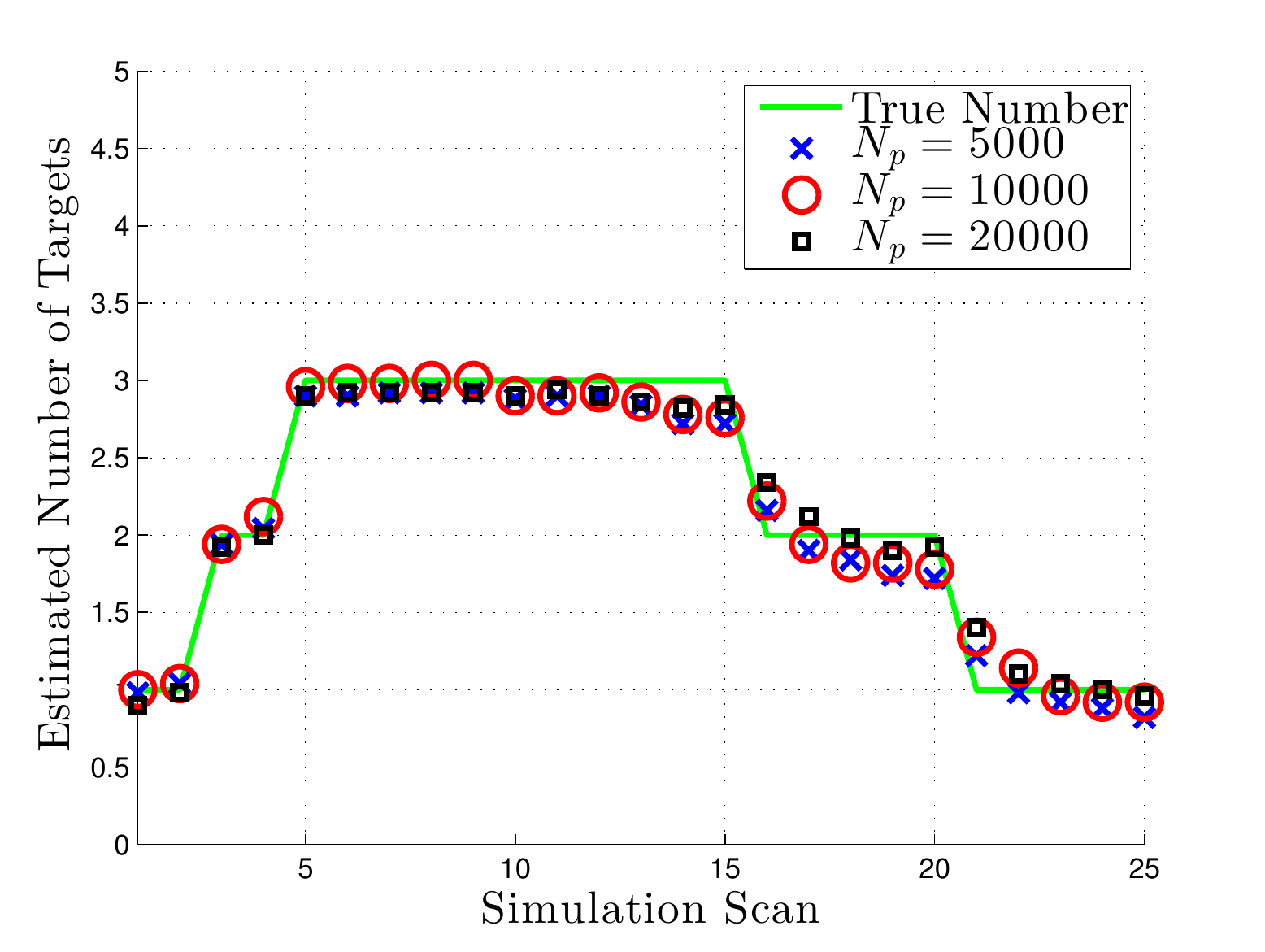}
	\label{fig:nt_7dB}
}
\hspace{2mm}
\subfigure[Monte Carlo results: OSPA Distance for $SNR=7dB$]{
	\includegraphics[width=.4\linewidth]{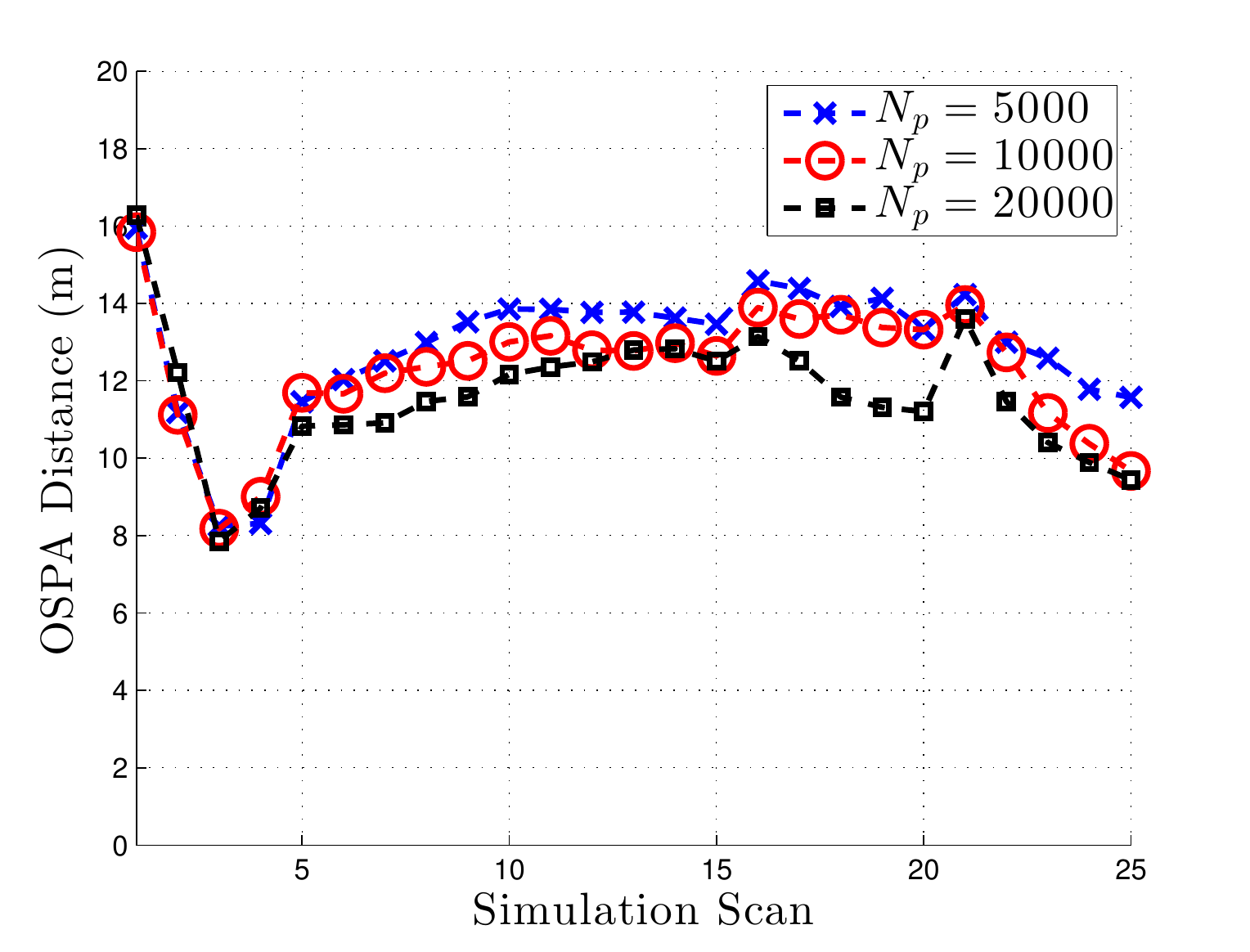}
	\label{fig:ospa_7dB}
}
\centering
\caption{Monte Carlo results for the case $SNR=7dB$}
\end{figure*}

Results in terms of the estimated number of targets and Optimal Sub-Pattern Assignment (OSPA) distance 
\cite{Schumacher08} are reported in Figs. \ref{fig:nt_10dB}-\ref{fig:ospa_10dB} for the case with $SNR=10dB$ 
and in Figs.  \ref{fig:nt_7dB}-\ref{fig:ospa_7dB} for the more difficult case with $SNR=7dB$. 
Notice from Fig. \ref{fig:ospa_10dB} the increase in OSPA distance after $k=3$ and $k=5$, i.e. the instants at 
which new targets enter the scene, and then reduces in time thanks to the filter convergence. 
For the more difficult case with $SNR=7dB$, we notice in Fig. \ref{fig:ospa_7dB} a slower filter convergence 
and increasing OSPA distance also for $k=15$ and $k=20$, i.e. the instants at which targets disappear. 
This is due to the fact that for lower $SNR$ the tracker prefers to retain a ``false'' track for few steps 
rather then declaring a ``dead'' target too soon, as confirmed by the time behaviour of the average estimated 
number of targets in Fig. \ref{fig:nt_7dB}. 
Tuning of the \textit{survival probability} $P_S$ can reduce this phenomenon. In practice, a perfect tuning of 
$P_S$ and of the \textit{birth probability} $P_B$ require additional prior knowledge on the surveillance area.
Overall, the results confirm the applicability of the proposed approach for challenging multi-target 
problems with closely spaced targets and low $SNR$. 


\section{Conclusions and Future Research}\label{sec:Conclusion}
In this paper we discussed a general solution for multi-target tracking with \textit{superpositional measurements}.
The proposed approach aims at evaluating the multi-target Bayes filter using SMC methods. 
The critical enabling step was the definition of an efficient proposal distribution based on the 
Approximate CPHD filter for \textit{superpositional 
measurements}. Numerical results confirmed the applicability to challenging multi-target tracking problems for
closely spaced targets using radar measurements with low SNR. 
A large-scale application of this approach might not be possible due to worsening depletion problems in high dimensional state spaces. 
However, MCMC methods could be used to devise a particle implementation that can scale with an increasing number of targets.
Furthermore, subdividing the targets into statistically non-interacting clusters and then processing the clusters separately
could lead to satisfactory performance with reduced computational load.
Finally, the capability of separating closely spaced targets for superpositional measurements, i.e. estimating the correct cardinality when 
there are unresolved targets, means the approach could also be used as an initialization block of cheaper trackers like the LMB 
\cite{Reuter2013} and Vo-Vo \cite{BTV13} filters.
Specifically, parts of the radar superpositional measurement could be processed with the proposed approach to 
find the correct number of targets as well as there location, then thresholded measurements could be processed with 
the LMB/Vo-Vo tracker. This should lead to improved performance as the LMB/Vo-Vo tracker would be using a more 
informative prior distribution.

\bibliographystyle{IEEEbib}

\begin{IEEEbiography}
[{\includegraphics[width=1in,height=1.25in,clip,keepaspectratio]{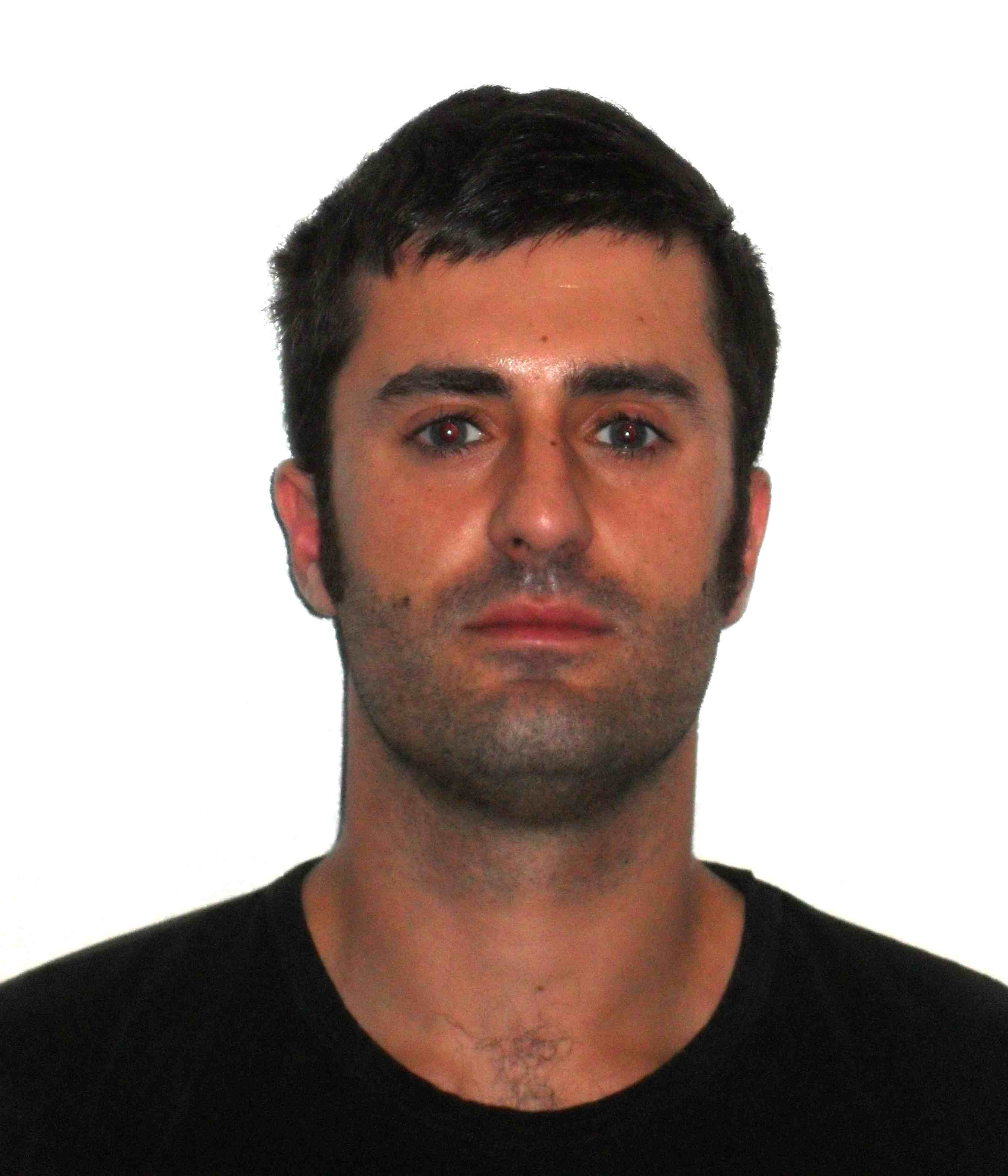}}]{Francesco Papi}
received the Laurea degree in Control Engineering in 2007 and a Ph.D. degree in
Computer Science and Control Engineering in 2011, both from the University of Firenze, Italy.
In January 2011 he joined Thales Nederland B.V. in Hengelo, the Netherlands, as a Marie Curie Research Fellow.
From January 2013 to April 2014 he was a Research Fellow at the Joint Research Centre, IPSC, European Commission,  Ispra, Italy.
He is currently Research Associate at the Department of Electrical and Computer Engineering, Curtin University, Perth, Australia.
His research interests include linear and nonlinear Bayesian estimation, single and multi-sensor target tracking, and data fusion.
\end{IEEEbiography}

\begin{IEEEbiography}
[{\includegraphics[width=1in,height=1.25in,clip,keepaspectratio]{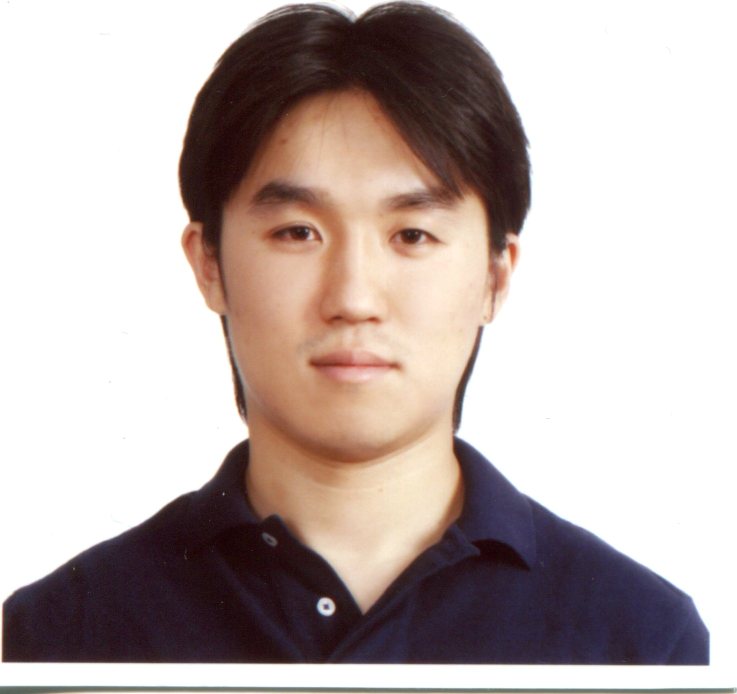}}]{Du Yong Kim}
received his B. E. degree in Electrical and Electronics Engineering from Ajou University, Korea in 2005. He received his M. S. and Ph. D. degrees from the Gwangju Institute of Science and Technology, Korea in 2006 and 2011, respectively. As a Postdoctoral researcher he has worked on statistical signal processing and image processing at the Gwangju Institute of Science and Technology, 2011-2012, University of Western Australia, 2012-2014. He is currently working as a research associate at the Department of Electrical and Computer Engineering, Curtin University. His main research interests include Bayesian filtering theory and its applications to machine learning, computer vision, sensor networks and automatic control.
\end{IEEEbiography}

\end{document}